\documentclass[]{aa}
\usepackage{graphicx}
\usepackage{natbib}
\usepackage{subfigure,xtab,booktabs}

\usepackage{siunitx}

\begin{document}

\title{A census of the pulsar population observed with the international LOFAR station FR606 at low frequencies (25-80~MHz)}

\author{L. Bondonneau
          \inst{1}
          \and
          J.-M. Grie{\ss}meier
          \inst{1,2}
     \and
          G. Theureau
          \inst{1,2,3}
          \and
          A. V. Bilous \inst{4}
          \and
          V. I. Kondratiev \inst{5,6}
          \and
          M. Serylak \inst{7,8}
          \and
          M. J. Keith
          \inst{9}
          \and
          A. G. Lyne
          \inst{9}
 }

\date{Version of \today}

\institute{
        LPC2E - Universit\'{e} d'Orl\'{e}ans / CNRS, France. \email{louis.bondonneau@cnrs-orleans.fr}
        \and
Station de Radioastronomie de Nan\c{c}ay, Observatoire de Paris - CNRS/INSU, USR 
  704 - Univ. Orl\'{e}ans, OSUC, route de Souesmes, 18330 Nan\c{c}ay, France
        \and
        Laboratoire Univers et Th\'eories LUTh, Observatoire de Paris, CNRS/INSU, Universit\'e Paris
Diderot, 5 place Jules Janssen, 92190 Meudon, France
        \and
        Anton Pannekoek Institute for Astronomy, University of Amsterdam, Science Park 904, 1098 XH Amsterdam, The Netherlands
        \and
        ASTRON, the Netherlands Institute for Radio Astronomy, Postbus 2, 7990 AA Dwingeloo, The Netherlands
        \and
        Astro Space Center, Lebedev Physical Institute, Russian Academy of Sciences, Profsoyuznaya Str. 84/32, Moscow 117977, Russia
        \and
        South African Radio Astronomy Observatory, 2 Fir Street, Black River Park, Observatory, Cape Town 7925, South Africa
        \and
    Department of Physics and Astronomy, University of the Western Cape, Cape Town 7535, South Africa
    \and
        Jodrell Bank Centre for Astrophysics, Department of Physics and Astronomy, The University of Manchester, Alan Turing Building, Oxford Road, Manchester, M13 9PL, UK
}

\date{Version of \today}

\abstract{
   To date, only 69 pulsars have been identified with a detected pulsed radio emission below 100 MHz.
   A LOFAR-core LBA census and a dedicated campaign with the Nan\c{c}ay LOFAR station in stand-alone mode were carried out in the years 2014$-$2017 in order to extend the known population in this frequency range.
   }
   {In this paper, we aim to extend the sample of known radio pulsars at low frequencies and to produce a catalogue in the frequency range of 25-80 MHz. 
   This will allow future studies to probe the local Galactic pulsar population,
   in addition to helping explain their emission mechanism,
   better characterising the low-frequency turnover in their spectra, 
   and obtaining new information about the interstellar medium through the study of dispersion, scattering, and scintillation.
   }
   { 
   We observed 102 pulsars that are known to emit radio pulses below 200 MHz and with declination above \SI{-30}{\degree}. We used the 
   the Low Band Antennas (LBA) of the LOw Frequency ARray (LOFAR) international station FR606 at the Nan\c{c}ay Radio Observatory in stand-alone mode, recording data between 25-80 MHz.
   }
   {Out of our sample of 102 pulsars, we detected 64. 
   We confirmed the existence of ten pulsars detected below 100 MHz by the LOFAR LBA census for the first time  \citep{bilous_lofar_2019} and we added 
   two more pulsars that had never before been detected in this frequency range. 
   We provided average pulse profiles, DM values, and mean flux densities (or upper limits in the case of non-detections).
   The comparison with previously published results allows us to identify a
   hitherto unknown spectral turnover for five pulsars, confirming the expectation that spectral turnovers are a widespread phenomenon.}
   {}


\titlerunning{A census of the pulsar population at low frequencies (25-80~MHz)}
\authorrunning{L. Bondonneau et al.}

\maketitle

\section{Introduction}



Until recently, radio frequencies below 100 MHz were largely under-explored in pulsar astronomy. The reasons for this are manifold: the interstellar medium causes high dispersion delays, which lead to pulse smearing unless coherent de-dispersion is used (computationally very expensive at such low frequencies); scattering on the inhomogeneities in the interstellar medium, leading to pulse smearing (regardless of the de-dispersion method); spectral turnover leading to low flux densities; the steep spectrum of the galactic background further reducing the measured signal-to-noise ratio (S/N); and the terrestrial ionosphere introducing angular shifts. Moreover, the times of arrival of pulsations extracted at such frequencies are highly affected by the profile frequency evolution due to the dependency of the emission altitude in the pulsar magnetosphere on the emission frequency 
\citep[radius-to-frequency-mapping; see, e.g.][]{ruderman_theory_1975,Cordes78}.

However, these effects do not only pose problems for observations. They also constitute a treasure trove of rich and complex phenomena which can be studied with sufficiently sensitive radio telescopes. 
For example, following the 
radius-to-frequency-mapping,
low-frequency radio emission traces the higher altitudes in the pulsars magnetosphere. As a consequence, a detailed wide-band study of low-frequency radio emission allows us, therefore, to map a large volume-fraction of the pulsar's magnetosphere.
Using observations with a large fractional bandwidth and high sensitivity at low frequencies, \citet[][]{hassall_wide-band_2012} were able to put strong constraints on the height of radio emission. 
Similarly, the precise measurement of the spectral turnover 
\citep[for which the physical cause is still unknown; see, e.g.][their Section 5]{bilous_lofar_2019}
will allow us to gain a better understanding of the pulsars' radio emission mechanism. 
Finally, temporal variations of the dispersion measure and of scattering can be monitored with very high precision to study the distribution of ionised plasma in the interstellar medium.\\


At the time of this writing, 2702 pulsars have been listed in the Australia Telescope National Facility (ATNF) Pulsar Catalogue\footnote{\url{http://www.atnf.csiro.au/research/pulsar/psrcat}, catalogue version 1.60} \citep{Manchester05}. Out of this population, 158 slow pulsars and 48 millisecond pulsars have been detected using the LOFAR core in the frequency range 110-188 MHz \citep[LOFAR HBA range,][]{bilous_lofar_2016,kondratiev_lofar_2016}. 

At frequencies below 100~MHz, the number of pulsars detected via their periodic, pulsed radio emission is considerably lower: 
40 pulsars have been detected by UTR-2 \citep{zakharenko_detection_2013}; 44 by LWA \citep{dowell_detection_2013,stovall_pulsar_2015}; 28 non-recycled pulsars and 3 millisecond pulsars by LOFAR-LBA \citep{pilia_wide-band_2016,kondratiev_lofar_2016}; and 2 millisecond pulsars by MWA \citep{bhat_observations_2018}.
Two additional pulsars have been previously reported at low significance (<5$\sigma$) by  \citet[][]{Reyes80} and \citet[][]{Deshpande92}, and three additional pulsars have been reported by \citet[][without pulse profiles]{Izvekova81}.
Combining these published results leads to a total of 69 different pulsars.
In a companion study \citep{bilous_lofar_2019}, we present the results of the LOFAR core LBA census, which contributes 14 pulsars which had not previously been detected at frequencies below 100 MHz.

Taken altogether, 83 different pulsars have been detected below 100 MHz prior to this study, 82 of which are located in the visible part of the sky as observed from the French LOFAR station (FR606).
This represents less than 20\% of the population of low-DM, non-recycled radio pulsars visible for the LOFAR station FR606. 

In view of the low number of pulsars known at frequencies below 100 MHz, we used the LOFAR station FR606 to conduct a systematic survey of the pulsar population below 100 MHz. Preliminary results of this survey have been already presented in \citet{griesmeier_interstellar_2018}.
The survey is now complete and this article details the final results. 

\section{Observations}\label{observation_sec}

Our observations were carried out with the International LOFAR Station in Nan\c{c}ay, FR606, used in stand\-alone mode, between 2016 and 2017. LOFAR, the Low Frequency Array, is fully described in \citet{stappers_observing_2011} and \citet{van_haarlem_lofar:_2013}.
The international LOFAR station FR606 contains 96 LBA dipoles. 
These antennas can operate over the range 10-90 MHz, with a central frequency of $\sim$50 MHz and a total bandwidth of up to 80 MHz. 
LOFAR is a digital telescope: Signals from individual LBA antennas are coherently summed, synthesizing a tied-array beam. In this study, we recorded data from 25-80 MHz (i.e.~a total bandwidth of 55 MHz) for pulsars with a DM < 17 pc cm$^{-3}$ and data from 50-80 MHz (i.e.~a total bandwidth of 30 MHz) for pulsars with higher DMs.

While a single LOFAR station such as FR606 only has a limited effective area, 
it allows us to take advantage of very flexible scheduling, especially 
for long observations or high cadence monitoring. The capability of this setup for pulsar science has already been demonstrated \citep{Rajwade16,Mereghetti16,Mereghetti17IAU,griesmeier_interstellar_2018, bondonneau_low_2018, tiburzi_usefulness_2019, michilli_low-frequency_2018,michilli_evolution_2018, Hermsen18,donner_first_2019}.\\ 

The sources of pulsating radio emission observed during our study were selected considering the pulsars previously detected at low frequencies by \citet{zakharenko_detection_2013} and \citet{stovall_pulsar_2015}.
We added some of the pulsars detected in the LOFAR HBA census \citep[$110-188$~MHz,][]{bilous_lofar_2016}, along with some additional pulsars we deemed interesting. We only kept radio sources with declination $\ge$\SI{-20}{\degree}. With this limit, the minimum elevation at meridian observed at  Nan\c{c}ay Radio Observatory is \SI{20}{\degree},
        and the effective area of the telescope is $\sim$11.5\% of the value for an observation at zenith. As an exception to this limit, we observed 
        the bright sources B0628-28 and B1749-28 down to an elevation of \SI{14}{\degree}. We discarded all pulsars with a dispersion measure higher than 140 pc cm$^{-3}$. Based on these criteria, we were left with 102 radio sources, as detailed in Table~\ref{tab:flux} (detections) and Table~\ref{tab:param:nondetected} (non-detections).

All the pulsars in the sample were observed for a duration from one to six hours, depending on the source elevation and on constraints related to the scheduling of the radio telescope.
Non-detections are based on observations of at least three hours. 
As a whole, the telescope time allocated to this project amounted to 294 hours (on average $\sim$3 h per pulsar).

\section{Data processing}

\subsection{Initial pulsar processing}

The nominal observing band (26-98 MHz) was split into three bands of 24 MHz each in order to spread the processing over three different computing nodes. 

        
To optimise the observing time, waveform data were systematically post-processed off-line when the radio telescope was pre-empted for observations in the International LOFAR Telescope (ILT) mode. Our pulsar processing pipeline was based on \texttt{DSPSR}\footnote{\texttt{https://github.com/demorest/dspsr}} \citep{van_straten_dspsr:_2011} which coherently de-dispersed
the data, folded the resulting time series at the period of the pulsar, and created sub-integrations of 10 seconds. Subsequently, observations were written out in \texttt{PSRCHIVE}\footnote{\texttt{http://psrchive.sourceforge.net/current/}} \citep{hotan_psrchive_2004} format.
After this step, the data from the three recording machines were combined into a single file.

Before the final analysis, each observation was refolded with an up-to-date ephemeris file when available \citep[compiled by][]{smith_searching_2019}. 
For 29 of the observed pulsars, we were able to use ephemeris files produced by the Jodrell Bank Observatory and the Nan\c{c}ay Radio Observatory. Refolded with a strong period accuracy, these observations are identified in Tables~\ref{tab:flux} and~\ref{tab:param:nondetected} by $^{\epsilon}$. In this case, it is no longer useful to search for period drifting. Consequently the search range is only in dispersion. 

Of these 29 ephemerides, most of them result from the timing analysis of observations made using the Lovell Telescope at Jodrell Bank~\citep[ongoing analysis carried out as a follow-up to][]{hobbs_2004}. The exceptions are the ephemeris of J2043$+$2740 and J2145-0750, which resulted from the timing analysis of the observations made using the Nan\c{c}ay radio-telescope by Ismaël~Cognard and Lucas~Gillemot (private communication); for details, see~\cite{cognard_2011}.

The dispersion measure (DM) values were provided by previous low-frequency observations \citep[mostly][]{zakharenko_detection_2013,bilous_lofar_2016}.

\subsection{Radio interference mitigation}

We used a custom radio frequency interference (RFI) mitigation scheme in order to automatically clean the observations. A few frequency channels near the top of the band, which was frequently polluted by radio transmission, were weighted to zero to improve the mitigation process.
RFI mitigation at such low frequencies is a challenge, and it is further complicated by the strongly peaked response of the LBA antennas
\citep[sensitivity maximum at $\sim$58 MHz, see][]{van_haarlem_lofar:_2013}. 
With a classical RFI mitigation technique (searching signal above a certain threshold), 
strong RFI signals in the low-sensitivity zone would be under-evaluated and not completely mitigated. To correct for this effect, each observation was (temporarily) flattened along the frequency axis by its (time-)average,  removing the frequency response of the instrument.
A mitigation mask was then generated by running \texttt{Coast Guard}\footnote{\texttt{https://github.com/plazar/coast\_guard/}} \citep{lazarus_prospects_2016} on this flattened dataset.
Finally, this mask was applied to the initial (un-flattened) datafile.

\subsection{Fine-tuning of DM and period }
\label{fine-tuning}
After RFI mitigation, we refined the pulsar's period and dispersion measure (DM) using \texttt{pdmp} (part of the software package \texttt{PSRCHIVE}).
This was required to account for deviations of these values from those in the ephemeris files used during the observations (e.g.~due to the limited precision of these files or due to a variation among these parameters). Given our frequency range, this was especially critical for the DM, where a small deviation from the nominal value can smear the pulse profile considerably.

This small correction to the DM is incoherent and can, in principle, result in a broadening of the pulse profile, which is more pronounced at low frequency as $\Delta t \propto DM (f_1^{-2} - f_2^{-2})$. 
In our sample of detected pulsars, this incoherent de-dispersion broadening ($\frac{\Delta t}{P0}$) 
does not affect the profile shape by more than one bin (out of a total of 512 phase bins).

The search range in pulsar period allows us to detect a drift up to one bin in a single sub-integration of 10 seconds corresponding to the same profile broadening than the dispersion range.

\subsection{Classification}

After visual inspection, pulsars were either classified as detections or non-detections. 
A pulsar was classified as a detection if (a) it had a signal-noise-ratio greater than 5, 
(b) was visible over a large frequency band, and (c) was detected in $\gtrsim$30\% of all sub-integrations.\footnote{In some cases, this can exclude pulsars with a large nulling fraction, see Section \ref{sec-disc-nondetections}.}

In some cases, remaining low-level RFI made the analysis ambiguous. In those cases, this RFI was manually cleaned using \texttt{pazi} (from the \texttt{PSRCHIVE} software package), and a new cycle of \texttt{pdmp} and visual inspection was required.


\subsection{Flux densities of detected pulsars}

Before calibration, we removed all data above 80 MHz and reduced the time resolution of the observation, increasing the length of an sub-integration to 60 seconds. This allowed us to considerably decrease the processing time of the calibration.

The flux calibration software we used is described in \citet{kondratiev_lofar_2016}. 
It is based on the radiometer equation \citep{dicke_measurement_1946}, the Hamaker beam model \citep{hamaker_understanding_2006}, and the \texttt{mscorpol} package by Tobia Carozzi. It calculates, for each frequency channel, the antenna response for the LBA station FR606 as a function of the pointing direction.

The fraction of flagged antennas (i.e. antennas not used during a given observation) was low (on average 2\% for our observations).
Due to its low impact compared to the effect of scintillation, we ignored this factor in the flux calculation.

\subsection{Upper limits for non-detected pulsars}
\label{sec-upperlimit}

For non-detected pulsars, we defined $S_\text{lim}$ as the upper limit for the mean flux density, according to the following equation \citep[following][]{lorimer_handbook_2004}:

\begin{equation}
S_\text{lim} = \frac{S/N (T_\text{inst}+T_\text{sky}) }{G\sqrt{n_\text{p} t_\text{obs} \Delta F_\text{eff}}} \sqrt{\frac{W/P}{1-W/P}}
\label{equLIM}
\end{equation}

Here,
\begin{itemize}
\item  $S/N=5$ is the signal-to-noise ratio limit required for a detection; 
\item  $T_\text{inst}$ is the (frequency-dependent) instrument temperature 
        \citep[deduced from an observation of Cassiopeia A, see][]{wijnholds_situ_2011};
\item  $T_\text{sky}$ is the sky temperature interpolated from a sky map at 408 MHz \citep{Haslam82}, scaled to our frequencies using $f^{-2.55}$ \citep{Lawson87};
\item  $G$ is the the effective gain, which depends on the source elevation. For this, we use the Hamaker beam model \citep{hamaker_understanding_2006} and the \texttt{mscorpol} package;
\item  $n_\text{p}=2$ is the number of polarizations;
\item  $t_\text{obs}$ is the duration of the observation;
\item  $\Delta F_\text{eff}$ is the effective bandwidth after RFI-cleaning; 
\item  $W$ and $P$ are the width of the integrated profile
and the pulse period, respectively. We assume a duty-cycle of $W/P=0.1$, which is consistent with the profiles of the detected pulsars.
\end{itemize}

Between $\sim$35-75 MHz,
the sky temperature $T_\text{sky}$ (which is frequency- and direction-dependent) dominates over the instrument temperature $T_\text{inst}$.
For example, at 60~MHz, $T_\text{sky}$ is 2350 K for pointing directions away from the Galactic plane (Galactic longitude of \SI{0}{\degree}, Galactic latitude of \SI{90}{\degree}),
but rises to 8500 K in the Galactic plane (Galactic longitude of \SI{90}{\degree}, Galactic latitude of \SI{0}{\degree}) 
and can reach up to 50000 K in the direction of the Galactic centre (Galactic longitude of \SI{0}{\degree}, Galactic latitude of \SI{0}{\degree}). 
For comparison, $T_\text{inst}=140$ K at 53 MHz.

\begin{figure}[ht]
\begin{center}
\includegraphics[height=5cm]{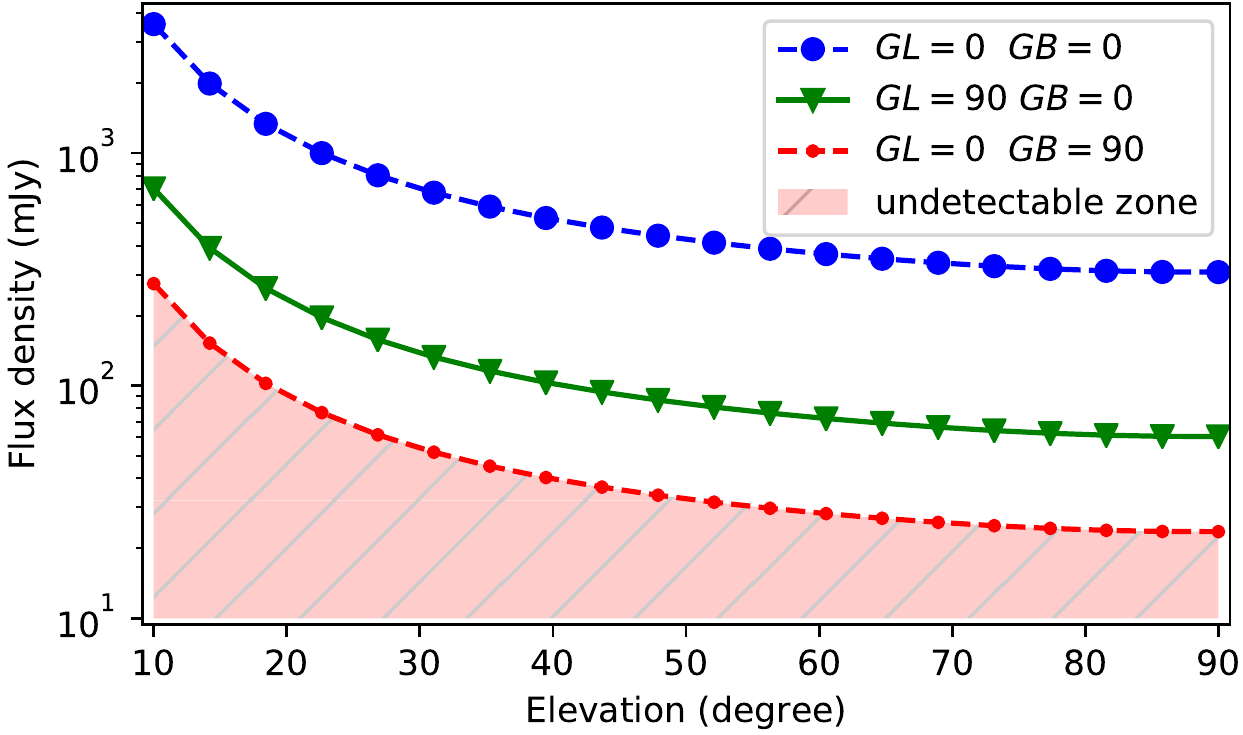}
\end{center}
\caption
{Minimum observable flux density depending on elevation 
and location of the radio source in the Galaxy for a S/N ratio of 5 and an observation duration of 4h. 
Three different Galactic temperature and location are used. 
Blue: Galactic centre ($gl=$\SI{0}{\degree}, $gb=$\SI{0}{\degree}, $T_\text{sky}=50000$ K); 
green: Galactic plane ($gl=$\SI{90}{\degree}, $gb=$\SI{0}{\degree}, $T_\text{sky}=8500$ K);
red: and outside the Galactic plane ($gl=$\SI{0}{\degree}, $gb=$\SI{90}{\degree}, $T_\text{sky}=2350$ K). 
Pulsars with flux densities in the colored region are too faint, and thus undetectable for the LBA antennas of the LOFAR station FR606.}
\label{limit_flux}
\end{figure}

Figure~\ref{limit_flux} shows the dependence of $S_\text{lim}$ 
on source elevation for three typical pointing directions 
(blue: towards the Galactic Centre, with $T_\text{sky}=50000$ K; 
green: in the Galactic Plane, with $T_\text{sky}=8500$ K;
red: outside the Galactic plane, with $T_\text{sky}=2350$ K).
The figure is based on Equation (\ref{equLIM}), and assumes an observation duration of $t_\text{obs}=4$h.
It shows that under optimal conditions (i.e.~a high elevation source outside the Galactic plane), we can achieve an upper flux limit of $\sim$30 mJy, whereas it can be up to three orders of magnitude less constraining for non-ideal conditions (low elevation source in the direction of the Galactic Centre).
Pulsars with mean flux densities in the colored region are not detectable, regardless of their position in the sky.

In Section \ref{sec-results}, we will use Equation (\ref{equLIM}) to derive 
upper limits 
for each of our non-detections on a case by case basis.

\section{Results}\label{sec-results}

\subsection{Detection rates}
\label{sec:res:rates}

Out of the 102 pulsating radio sources we observed, we successfully detected 64 pulsars (61 slow pulsars and 3 millisecond pulsars).
12 of these pulsars were either\ detected in this frequency range for the first time or 
were detected contemporaneously in this study and in the LOFAR core LBA census \citep[see companion article by][]{bilous_lofar_2019}. 
Most of these `new' low-frequency detections overlap with the simultaneous ana\-lysis of LOFAR core data 
\citep[10 out of 12, cf.][]{bilous_lofar_2019}. 
Compared to \citet{bilous_lofar_2019}, we detected two additional pulsars
(B0105$+$65 and B2021$+$51) that were not in their sample
and which were previously undetected at frequencies below 100 MHz. 

\begin{figure}[ht]
\begin{center}
\includegraphics[height=7cm]{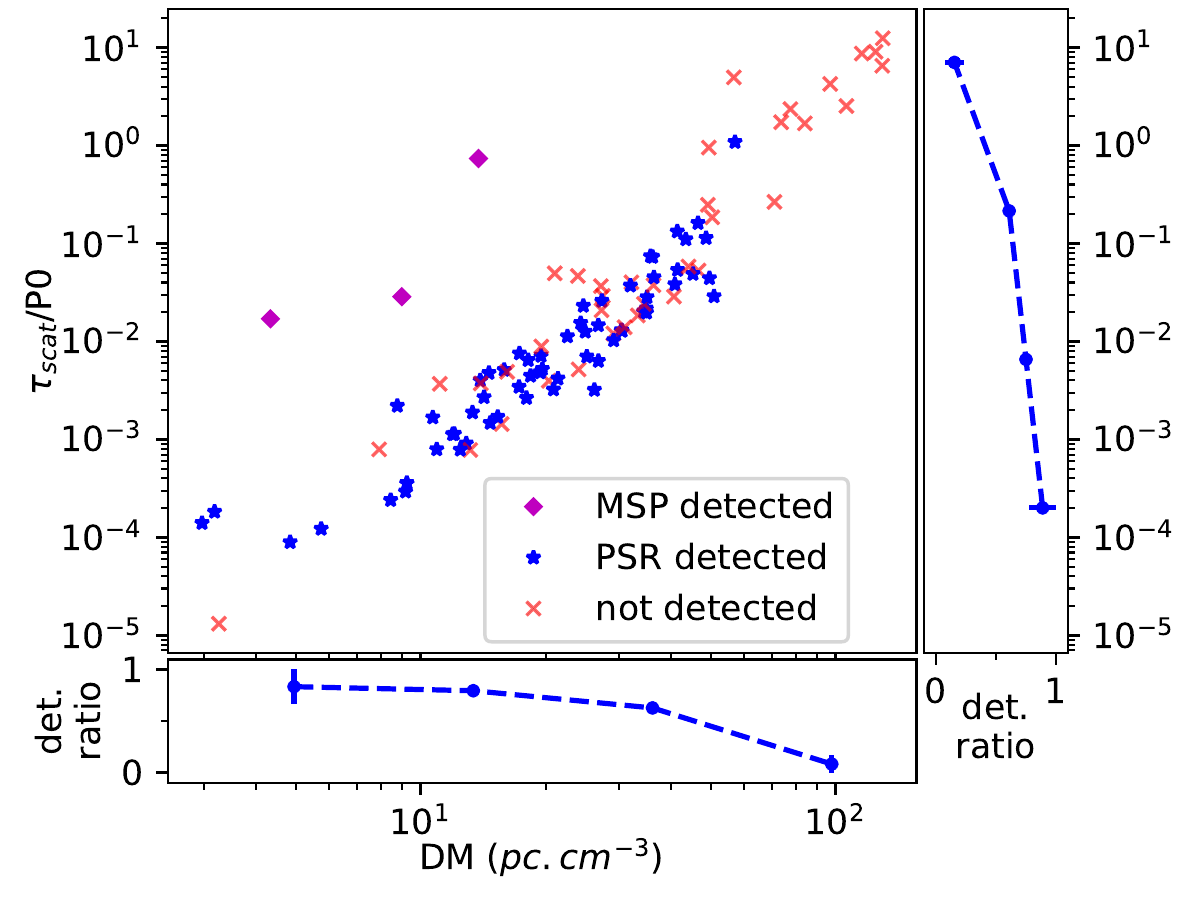}
\end{center}
\caption
{Scattering time in units of the pulsar period versus the dispersion measure for the pulsars of our sample (centre plot, double-logarithmic axes). Detected slow pulsars are shown as blue stars, millisecond pulsars as magenta diamonds, and non-detections as red crosses. Right and bottom panels (with semi-logarithmic axes) shows the fraction of detected pulsars for each axis of the central plot.}
\label{Scat_P0}
\end{figure}

Figure~\ref{Scat_P0} compares the detected pulsars (blue and magenta points) and the non-detections (red crosses) in terms of measured DM and expected scattering delay at 60 MHz \citep[calculated using the model of][]{yao_new_2017} 
relative to the pulsar period. 
The two small plots indicate the fraction of detected pulsars as a function of DM (lower panel) and scattering delay (right panel).
As expected, pulsars become undetectable once the scatter broadening exceeds the pulsar period (central plot, and right panel). 
The exception to this rule (B0355$+$54) is discussed in Section \ref{sec:disc:scattering}.

Figure~\ref{Scat_P0} also shows a correlation between high DM and high scattering timescales. This 
correlation is well-known, and has been described, for example,~by \citet{bhat_multifrequency_2004}.
This correlation allows us to estimate the maximum DM at which we can detect pulsars before their pulsations become undetectable due to scatter-broadening. 
In our case, the maximum detected DM value is $\sim57$ (for B0355$+$54).

\begin{figure}[ht]
\begin{center}
\includegraphics[height=9cm, trim=0.5cm 0cm 0cm 0cm]{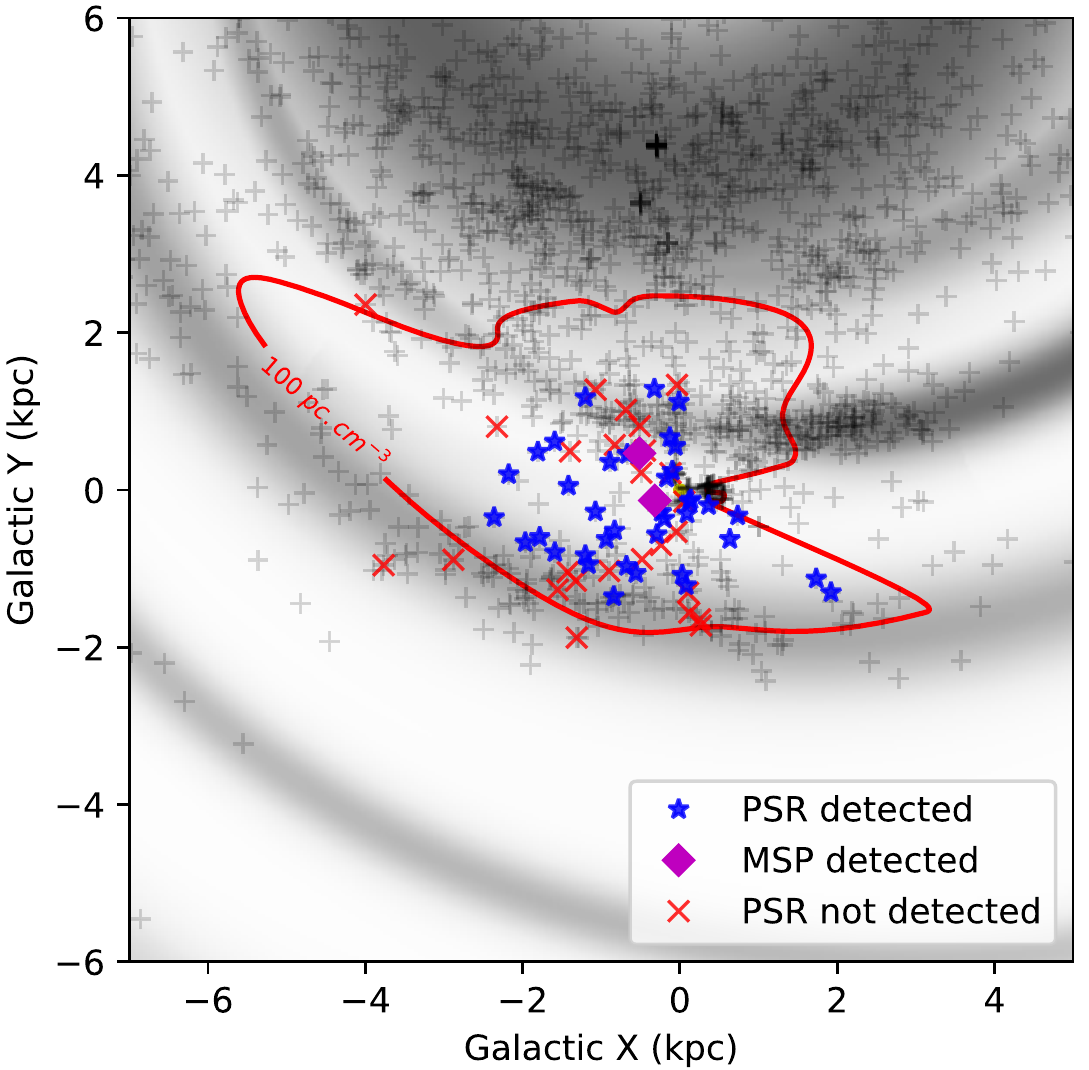}
\end{center}
\caption
{Representation of the census in the Galactic plane with the Earth at the origin of the axes. The electron density model YMW16 \citep{yao_new_2017} is represented in grey-scale. Blue stars: detected pulsars.  Magenta diamonds: detected MSPs. Red crosses: non-detected pulsars. Only pulsars in the galactic plane are plotted (distance to the galactic plane less than 700 pc). The red line is an isocontour for a DM of \SI{100}{pc\,cm^{-3}}. Pulsar distances are derived from the dispersion measure using YMW16 \citep{yao_new_2017}.}
\label{galactic_map}
\end{figure}

Of course, the DM is related to distance. We can, thus, estimate the maximum distance at which we can detect pulsars. 
For this, we 
look at the spatial distribution of our observations and detections. 
Figure~\ref{galactic_map} shows the location of our sources in the Galactic plane, with the Earth at the origin of the axes. 
The electron density model YMW16 from \citet{yao_new_2017} is represented in a gray scale.
Pulsars detected in the present survey are shown as blue
dots for normal pulsars and magenta diamonds for the MilliSecond Pulsars (MSPs), and non-detections are shown with red crosses. For this, pulsar distances are derived from the DM using the electron density model YMW16 \citep{yao_new_2017}. Only pulsars in the Galactic plane are shown (Galactic latitude between \ang{-20} and \ang{20}). 

A red isocontour denotes a dispersion measure of 100~pc\,cm$^{-3}$ in the Galactic plane ($gb$=\ang{0}), corresponding to 
a scattering time of one second at 60 MHz 
\citep[derived from the Galactic density model of][]{yao_new_2017}. 
With such a scattering delay, even the pulsations from slow pulsars are smeared out and become undetectable.
Indeed, we do not have any detection beyond this isocontour.
%
One can see that the red line is at a distance of only a few kpc of the Solar System. 
Indeed, low-frequency observations of pulsed signals are limited to the nearby population. 
For comparison, it is possible to observe sources close to the Galactic centre for observations at 1 GHz.


\subsection{Detected pulsars}
\label{sec-detections}

For each detected pulsar, we measured the 
spin period $P0$, the DM, and the mean flux density, and we calculated the expected scattering delay $\tau_\text{scat}^\text{calc}$ 
\citep[derived from the Galactic density model of][]{yao_new_2017}.
These values are detailed in Table~\ref{tab:flux}.



\begin{figure*}[ht]
\begin{center}
\includegraphics[height=11cm, trim=0cm 9cm 0cm 8cm]{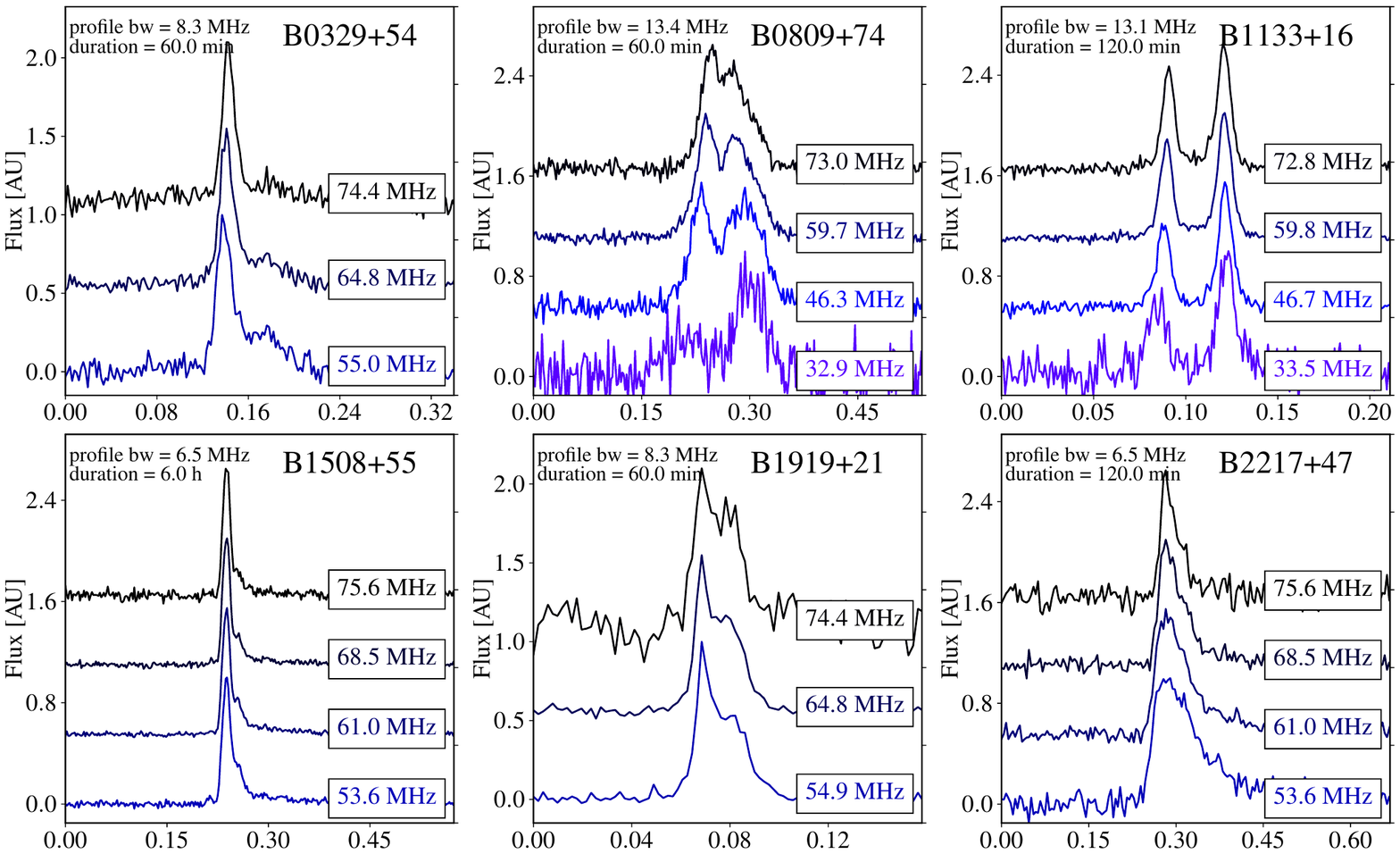}
\end{center}
\caption[freqevol]{Frequency-dependent profiles for the six pulsars with the best  S/N in our sample. Profiles are zoomed in on the on-pulse region.}
\label{freqevol}
\end{figure*}

For each detected pulsar, an 
average pulse
profile was generated. 
These profiles are shown in Figure~\ref{ALL:prof}. 
When the signal-to-noise ratio is sufficient, pulse profiles can be compared at  different observing frequencies. 
This allows us to reveal the frequency dependence of the beam pattern. This is illustrated in Figure~\ref{freqevol} for the six pulsars with the best signal-to-noise ratio in our sample:\\
B0329$+$54 is a mode-switching pulsar \citep[see, e.g.][]{Chen_2011}. In the long observation period (1h) shown in Figure~\ref{freqevol}, we observed a mix of both modes.\\
Pulsar B0809$+$74 shows drifting sub-pulses and has a frequency-dependent profile. It is discussed in detail in \citet{hassall_differential_2013}.\\
Pulsar B1133$+$16 shows two components where the separation, the amplitudes, and widths are frequency-dependent. This frequency dependence of the beam pattern is visible on Figure~\ref{freqevol} and in  \citet{hassall_wide-band_2012}.\\
B1508$+$55 shows pulse echoes which drift across the profile on the timescale of a few years \citep[see][]{Wucknitz_2019}. This is the reason why the profile in Figure~\ref{freqevol} is different than that in \citet{bilous_lofar_2019}.\\
In contrast to the majority of pulsars detectable at low frequency, the separation between the components of B1919$+$21 decreases with decreasing frequency \citep[see][]{hassall_wide-band_2012}. In addition, the relative amplitudes of the components seem to be frequency-dependent.\\
Pulsar B2217$+$47 is highly affected by time-variable scattering \citep[][]{michilli_low-frequency_2018,donner_first_2019}. This effect, convolved with the profile, produces a frequency dependent exponential tail which is clearly visible in Figure~\ref{freqevol}.\\

A mean flux density value over the entire band was obtained for each detection (Table~\ref{tab:flux}). 
The measured flux density is assumed to be correct within a 50\% accuracy, as was recommended for LOFAR HBA observations \citep{bilous_lofar_2016,kondratiev_lofar_2016} based on the comparison of flux measurements with literature data.  
This uncertainty includes refractive scintillation, intrinsic flux variation, and imperfections in the beam model used for calibration \citep{kondratiev_lofar_2016}. According to our estimates, refractive scintillation represents the dominant effect 
(at 60 MHz, the timescale for refractive scintillation is of the order of at least one month).

We should note that this estimation was originally established for LOFAR observations in the HBA band (100-200 MHz); for observations with a single station at frequencies below 100 MHz, the situation might be slightly different. This will be studied in more detail in a future paper.
\onecolumn


\addtocounter{table}{-1}
\topcaption{top}
\tablecaption{Pulsars detected in this census. 
JName, BName: pulsar name. P0: pulsar period. 
DM: the best-fit DM calculated using \texttt{pdmp}. 
$\tau^\text{calc}_\text{scat}$/P0: the scattering time 
as estimated using YMW16, at 60 MHz divided by the pulsar period, expressed in \%. 
duty cycle: the effective width in pulses profiles (based on w50), expressed in \%.
S/N: the signal-to-noise-ratio of the detected pulsar profile.
duration: total observation length in minutes. 
{$f_{\text{centre}}$}: the centre frequency of the observation in MHz.
avg.elev: the average elevation during the  observation.
mean flux: the mean flux density determined for the corresponding centre frequency, 
with an error bar of 50\%.
$^{\sigma}$: pulsed flux density only (due to scatter broadening, part of the pulsar's energy reaches the telescope as continuum, see~Section \ref{sec:disc:scattering}).
$^{\tau}$: the dispersion measure is not corrected for the effect of scatter-broadening (see Section \ref{fiducial}).
$^{\epsilon}$: the dispersion measure is not corrected for the effect of intrinsic profile evolution with frequency (see Section \ref{fiducial}).
$^{\epsilon}$: the file is folded using an ephemeris file from either Jodrell Bank Observatory or Nan\c{c}ay Radio Observatory (see Section \ref{fine-tuning}).
\label{tab:flux}}
\tablefirsthead{\toprule \small{J2000} &\small{Discovery} &\small{P0} &\small{DM}                                               & \small{$\tau^\text{calc}_\text{scat}$/P0} & \small{duty} & \small{SNR} & \small{duration} & \small{$f_\text{centre}$}                    & \small{avg.elev}      & \small{mean}  \\
 \small{Name}&\small{Name}& \small{[s]} & \small{[pc cm$^{-3}$]}                        & \small{[\%]}  & \small{cycle [\%]}&  & \small{[min]}  & \small{[MHz]}                   &       \small{degrees} & \small{flux [mJy]} \\ \midrule}
\tablehead{%
\multicolumn{3}{c}%
{{\bfseries  Continued from previous page}} \\
\toprule \small{J2000} &\small{Discovery} &\small{P0} &\small{DM}                                               & \small{$\tau^\text{calc}_\text{scat}$/P0} & \small{duty} & \small{SNR} & \small{duration} & \small{$f_\text{centre}$}                    & \small{avg.elev}      & \small{mean}  \\
 \small{Name}&\small{Name}& \small{[s]} & \small{[pc cm$^{-3}$]}                        & \small{[\%]}  & \small{cycle [\%]}&  & \small{[min]}  & \small{[MHz]}                   &       \small{degrees} & \small{flux [mJy]} \\ \midrule}
\tabletail{%
\midrule \multicolumn{3}{r}{{Continued on next page}} \\ \midrule}
\tablelasttail{%
\multicolumn{3}{r}{{}} \\ \bottomrule}
\begin{xtabular*}{\textwidth}{llllllllllll}
J0014$+$4746 & B0011$+$47 & 1.241$^{\epsilon}$ & 30.30(2) & 1.3 & 4.4 & 9 & 240 & 65 & 79 & 43(21) \\
J0030$+$0451 & J0030$+$0451 & 0.005 & 4.33320(6) & 1.7 & 4.9 & 9 & 180 & 53 & 46 & 86(43) \\
J0034$-$0534 & J0034$-$0534 & 0.002 & 13.76580(4)$^{\tau}$ & 73.8 & 46.6 & 55 & 180 & 53 & 32 & 855(428) \\
J0034$-$0721 & B0031$-$07 & 0.943 & 10.916(5) & 0.1 & 15.3 & 41 & 120 & 53 & 35 & 560(280) \\
J0051$+$0423 & J0051$+$0423 & 0.355 & 13.9265(5) & 0.4 & 5.3 & 14 & 120 & 53 & 46 & 30(15) \\
J0056$+$4756 & B0053$+$47 & 0.472 & 18.14(1) & 0.6 & 6.5 & 17 & 135 & 65 & 84 & 102(51) \\
J0108$+$6608 & B0105$+$65 & 1.284 & 30.56(2) & 1.3 & 4.0 & 11 & 325 & 65 & 67 & 74(37) \\
J0141$+$6009 & B0138$+$59 & 1.223 & 34.931(2) & 2.2 & 10.8 & 28 & 120 & 65 & 68 & 242(121) \\
J0152$-$1637 & B0149$-$16 & 0.833 & 11.9289(5) & 0.1 & 10.4 & 37 & 120 & 53 & 25 & 215(107) \\
J0323$+$3944 & B0320$+$39 & 3.032 & 26.20(1) & 0.3 & 20.8 & 31 & 165 & 65 & 73 & 127(63) \\
J0332$+$5434 & B0329$+$54 & 0.715 & 26.768(1) & 1.5 & 11.2 & 119 & 60 & 65 & 51 & 1841(921) \\
J0358$+$5413 & B0355$+$54 & 0.156 & 57.15(1)$^{\tau}$ & 109 & 7.9 & 9 & 240 & 65 & 77 & 129(64)$^{\sigma}$ \\
J0454$+$5543 & B0450$+$55 & 0.341 & 14.5921(10) & 0.5 & 8.7 & 28 & 325 & 53 & 53 & 124(62) \\
J0528$+$2200 & B0525$+$21 & 3.746 & 50.90(2)$^{\epsilon}$ & 2.9 & 13.7 & 31 & 120 & 53 & 61 & 257(128) \\
J0611$+$30 & J0611$+$30 & 1.412 & 45.31(4) & 4.9 & 4.5 & 9 & 240 & 65 & 68 & 64(32) \\
J0630$-$2834 & B0628$-$28 & 1.244 & 34.42(1) & 2.0 & 7.9 & 23 & 55 & 65 & 14 & 1076(538) \\
J0700$+$6418 & B0655$+$64 & 0.196$^{\epsilon}$ & 8.7749(2) & 0.2 & 10.7 & 33 & 120 & 53 & 68 & 95(47) \\
J0814$+$7429 & B0809$+$74 & 1.292 & 5.7578(9)$^{\epsilon}$ & 0.0 & 45.9 & 252 & 60 & 53 & 62 & 1630(815) \\
J0820$-$1350 & B0818$-$13 & 1.238 & 40.962(10) & 3.8 & 2.7 & 10 & 115 & 65 & 28 & 61(31) \\
J0826$+$2637 & B0823$+$26 & 0.531 & 19.4743(8) & 0.7 & 14.2 & 79 & 60 & 65 & 63 & 423(212) \\
J0837$+$0610 & B0834$+$06 & 1.274 & 12.864(2)$^{\epsilon}$ & 0.1 & 6.9 & 309 & 60 & 65 & 44 & 1268(634) \\
J0908$-$1739 & B0906$-$17 & 0.402 & 15.875(2) & 0.5 & 1.5 & 5 & 180 & 53 & 24 & 29(15) \\
J0922$+$0638 & B0919$+$06 & 0.431 & 27.2965(5) & 2.6 & 15.4 & 144 & 180 & 65 & 42 & 550(275) \\
J0927$+$23   & J0927$+$23 & 0.762 & 23.127(2) & 0.8 & 0.5 & 6 & 215 & 62 & 54 & 12(6) \\
J0946$+$0951 & B0943$+$10 & 1.098 & 15.3291(5) & 0.2 & 15.2 & 148 & 150 & 53 & 46 & 610(305) \\
J0953$+$0755 & B0950$+$08 & 0.253 & 2.9711(2)$^{\epsilon}$ & 0.0 & 14.6 & 140 & 60 & 62 & 41 & 2276(1138) \\
J1115$+$5030 & B1112$+$50 & 1.656 & 9.197(3) & 0.0 & 2.5 & 12 & 275 & 53 & 75 & 21(11) \\
J1136$+$1551 & B1133$+$16 & 1.188 & 4.8468(7)$^{\epsilon}$ & 0.0 & 18.7 & 261 & 120 & 53 & 53 & 894(447) \\
J1238$+$2152 & J1238$+$2152 & 1.119 & 17.967(3) & 0.3 & 4.1 & 15 & 155 & 65 & 62 & 38(19) \\
J1239$+$2453 & B1237$+$25 & 1.382 & 9.2562(8)$^{\epsilon}$ & 0.0 & 0.2 & 50 & 170 & 62 & 65 & 102(51) \\
J1313$+$0931 & J1313$+$0931 & 0.849 & 12.0318(5) & 0.1 & 2.2 & 7 & 215 & 57 & 48 & 25(13) \\
J1321$+$8323 & B1322$+$83 & 0.670$^{\epsilon}$ & 13.28(3) & 0.2 & 1.4 & 5 & 225 & 53 & 43 & 16(8) \\
J1509$+$5531 & B1508$+$55 & 0.740 & 19.616(1) & 0.5 & 10.7 & 378 & 360 & 65 & 73 & 943(471) \\
J1532$+$2745 & B1530$+$27 & 1.125 & 14.697(6) & 0.1 & 6.0 & 18 & 240 & 53 & 66 & 69(35) \\
J1543$-$0620 & B1540$-$06 & 0.709 & 18.372(4) & 0.4 & 9.0 & 22 & 145 & 65 & 34 & 143(72) \\
J1543$+$0929 & B1541$+$09 & 0.748 & 34.950(5) & 3.5 & 5.6 & 26 & 90 & 53 & 50 & 541(270) \\
J1607$-$0032 & B1604$-$00 & 0.422 & 10.6823(5) & 0.2 & 9.9 & 64 & 60 & 53 & 42 & 575(288) \\
J1614$+$0737 & B1612$+$07 & 1.207 & 21.401(2) & 0.4 & 3.8 & 24 & 165 & 65 & 48 & 120(60) \\
J1635$+$2418 & B1633$+$24 & 0.491$^{\epsilon}$ & 24.24(1) & 1.5 & 3.9 & 8 & 210 & 65 & 56 & 68(34) \\
J1645$-$0317 & B1642$-$03 & 0.388 & 35.7589(5) & 7.4 & 8.8 & 48 & 240 & 65 & 37 & 420(210) \\
J1645$+$1012 & J1645$+$1012 & 0.411 & 36.165(6) & 7.3 & 4.6 & 12 & 165 & 65 & 50 & 108(54) \\
J1709$-$1640 & B1706$-$16 & 0.653 & 24.892(2) & 1.3 & 7.3 & 17 & 120 & 53 & 25 & 317(159) \\
J1740$+$1311 & B1737$+$13 & 0.803 & 48.660(5)$^{\tau}$ & 11.3 & 4.6 & 14 & 180 & 65 & 36 & 131(66) \\
J1741$+$2758 & J1741$+$2758 & 1.361 & 29.168(8) & 1.0 & 9.6 & 21 & 210 & 65 & 67 & 54(27) \\
J1758$+$3030 & J1758$+$3030 & 0.947 & 35.08(1) & 2.8 & 2.0 & 8 & 115 & 65 & 67 & 26(13) \\
J1813$+$4013 & B1811$+$40 & 0.931 & 41.60(2) & 5.4 & 2.1 & 11 & 115 & 65 & 57 & 68(34) \\
J1825$-$0935 & B1822$-$09 & 0.769 & 19.386(2) & 0.5 & 5.7 & 33 & 120 & 53 & 32 & 2502(1251) \\
J1840$+$5640 & B1839$+$56 & 1.653 & 26.773(2) & 0.6 & 7.0 & 166 & 180 & 65 & 57 & 481(240) \\
J1844$+$1454 & B1842$+$14 & 0.375 & 41.483(2)$^{\tau}$ & 13.2 & 6.7 & 36 & 120 & 65 & 51 & 773(386) \\
J1921$+$2153 & B1919$+$21 & 1.337 & 12.437(2) & 0.1 & 8.4 & 180 & 60 & 65 & 63 & 1586(793) \\
J1932$+$1059 & B1929$+$10 & 0.227 & 3.186(2) & 0.0 & 5.8 & 15 & 120 & 53 & 41 & 306(153) \\
J1955$+$5059 & B1953$+$50 & 0.519 & 31.990(5) & 3.7 & 1.2 & 5 & 225 & 65 & 70 & 16(8) \\
J2018$+$2839 & B2016$+$28 & 0.558 & 14.1982(5) & 0.3 & 8.6 & 37 & 135 & 53 & 54 & 243(121) \\
J2022$+$2854 & B2020$+$28 & 0.343 & 24.6315(10) & 2.3 & 6.2 & 30 & 150 & 65 & 69 & 243(122) \\
J2022$+$5154 & B2021$+$51 & 0.529 & 22.541(6) & 1.1 & 1.3 & 6 & 120 & 65 & 81 & 46(23) \\
J2113$+$2754 & B2110$+$27 & 1.203 & 25.121(3) & 0.7 & 8.4 & 37 & 150 & 65 & 65 & 143(71) \\
J2145$-$0750 & J2145$-$0750 & 0.016$^{\epsilon}$ & 9.0058(2) & 2.9 & 3.5 & 8 & 180 & 53 & 34 & 59(30) \\
J2219$+$4754 & B2217$+$47 & 0.538 & 43.492(1)$^{\tau}$ & 11.0 & 15.6 & 125 & 120 & 65 & 76 & 1239(620) \\
J2225$+$6535 & B2224$+$65 & 0.683 & 36.473(4) & 4.5 & 5.5 & 29 & 210 & 65 & 46 & 293(146) \\
J2305$+$3100 & B2303$+$30 & 1.576 & 49.60(3) & 6.2 & 4.1 & 11 & 115 & 65 & 72 & 45(22) \\
J2308$+$5547 & B2306$+$55 & 0.475$^{\epsilon}$ & 46.57(4)$^{\tau}$ & 16.2 & 2.3 & 5 & 120 & 65 & 70 & 170(85) \\
J2313$+$4253 & B2310$+$42 & 0.349 & 17.282(8) & 0.8 & 2.6 & 11 & 240 & 65 & 60 & 81(41) \\
J2317$+$2149 & B2315$+$21 & 1.445 & 20.876(5) & 0.3 & 3.7 & 10 & 120 & 65 & 59 & 35(18) \\
J2330$-$2005 & B2327$-$20 & 1.644 & 8.4554(10) & 0.0 & 4.0 & 16 & 120 & 53 & 22 & 111(55) \\
\end{xtabular*}

\addtocounter{table}{-1}
\twocolumn

\subsection{Upper limits for non-detections}

For each non-detection, 
we computed an upper limit for the mean flux density according to Equation (\ref{equLIM}) in Section \ref{sec-upperlimit}. 
The resulting values are given in Table~\ref{tab:param:nondetected}. Depending on source position and observation time, we obtained upper limits between $\sim60$ and 4500 mJy, which is compatible with our initial expectation (Figure \ref{limit_flux}).

Compared to previous observations, 
only 20 pulsars previously detected below 100 MHz have either not been observed or not been detected.
Of these, one (J0437$-$4715) is not observable for FR606 and seven others 
have not been observed as part of this survey. 
This leaves 13
previously reported pulsars which we have not detected, some of which were reported as faint or marginal detections:\\
    B0226$+$70 was detected by the LOFAR core LBA census      
        \citep{bilous_lofar_2019} with a mean flux density of 49 mJy, which is consistent with our upper limit of 329 mJy.\\
    B0301$+$19 has been reported by 
        \citet[][$\sim$40 mJy at 61 MHz]{Izvekova81}, by
        \citet[][120$\pm$60 mJy at 64.5 MHz]{stovall_pulsar_2015},
        and has been detected in the LOFAR core LBA census 
        \citep[][61$\pm$33 mJy at 50 MHz]{bilous_lofar_2019}.
        Our upper limit is 121 mJy, which is compatible with all of these detections.\\ 
    B0609$+$37 was detected by the LOFAR core LBA census      
        \citep{bilous_lofar_2019} with a mean flux density of 46 mJy, which is consistent with our upper limit of 664 mJy.\\
    B0611$+$22 and B0656$+$14 were previously reported as a detections with flux densities 
    \citep[][180 mJy and 60 mJy at 85 MHz respectively]{Izvekova81}, which are compatible with our upper limits of 337 mJy and 77 mJy.\\
    We expected to detect J0921$+$6254 
        \citep[detected in][but with no measured flux density]{pilia_wide-band_2016}.
        The pulsar was detected by the LOFAR core LBA census  
        \citep{bilous_lofar_2019} with a mean flux density of 41$\pm$22 mJy,  which is consistent with our upper limit of 58 mJy.\\
    B0940$+$16 was a weak detection in 
        \citet[][7.3 mJy at 25 MHz]{zakharenko_detection_2013}, which is compatible with our upper flux limit of 430 mJy (taking into account their spectral index of $\alpha=2.31$).\\
    B1749$-$28 has been reported twice 
        \citep{Izvekova81,stovall_pulsar_2015}, but is at very low elevation for FR606, which strongly reduced the efficiency of the antenna array. This is reflected in our poorly constraining upper limit of 4533 mJy, which is compatible with the previous detections.\\
    B1839$+$09 was detected by the LOFAR core LBA census  
        \citep{bilous_lofar_2019} with a mean flux density of 190 mJy, which is consistent with our upper limit of 521 mJy.\\
    J1851$-$0053 and J1908$+$0734 were weak detections in 
        \citet[][7 mJy for both pulsars at 25 MHz]{zakharenko_detection_2013}. 
        Our upper flux limit of 578 mJy and 203 mJy are compatible with their data and constrains the spectral index to values $<5$ and $< 4$ respectively.\\
        Based on \citet[][27 mJy at 25 MHz]{zakharenko_detection_2013}, we hoped to detect B1944$+$17. Still, their values are compatible with our 
        upper limit of 110 mJy
        (taking into account their spectral index of $\alpha=1.22$).\\
        Based on \citet[][15 mJy at 25 MHz]{zakharenko_detection_2013}, we hoped to detect B1952$+$29 for which we have an upper limit of 124 mJy. 
        Our non-detection is compatible with their data 
        and constrains the spectral index to values $< 2.5$.\\
Besides the lack of sensitivity, other possible reasons for non-detections are discussed in Section \ref{sec-disc-nondetections}.

\section{Discussion}

\subsection{Dispersion at low frequency}

In Section \ref{sec-detections}, we present DM values for all pulsars detected in this census.
To obtain these values, we used \texttt{pdmp,} which modifies the DM value until the S/N of the pulse profile is maximised.

This approach does not correctly take into account frequency-dependent pulse profile variations. A typical example for this would be a 
pulsar with two or more bright components, whose flux ratio changes as a function of frequency such as B1133$+$16 and B0809$+$74 (cf.~Figure \ref{freqevol}).

A similar situation arises for pulsars that are scatter-broadened. In that case, part of the scatter-broadening ($\propto$~$f^{-4}$) is absorbed by \texttt{pdmp}, resulting in an erroneous extra correction of the DM ($\propto$~$f^{-2}$), especially at low frequencies.



An ideal de-dispersion process should use a 2D template, either based on Gaussian fits \citep{Pennucci14,Liu14} or
based on the smoothed dataset \citep[e.g.][]{donner_first_2019}. 
In addition, a fiducial \label{fiducial} point would be required \citep[e.g.][]{hassall_wide-band_2012} 
in order to disentangle dispersion and frequency-dependent profile variation.
Without this, the absolute DM cannot be measured.

We did not apply any of these methods in this publication, which limits the DM precision for some of the pulsars in this census. These pulsars are clearly labelled in Table \ref{tab:flux}.

\subsection{Dispersion, scattering, and detection rate}
\label{sec:disc:scattering}

As discussed in Section \ref{sec:res:rates}, dispersion and scattering are correlated. Indeed, 
Figure~\ref{Scat_P0} shows that the measured DM and the calculated scattering time from YMW16 
are correlated in our sample of detected pulsars. The detection rates decrease for high scattering time and high DM.
Low-frequency observations are highly affected by the dispersion introduced by the interstellar medium. However, this effect is corrected using coherent de-dispersion \citep{hankins_pulsar_1975,BondonneauURSI}.

Consequently, the low detection rate in high DM is not due to the dispersion, but caused by the multi-path propagation in the interstellar medium which is usually described by a convolution between the pulse profile and an exponential function. The result is an exponential tail characterised by the scattering time $\tau$, as can be seen, for example, with B2217$+$47 in Figure~\ref{ALL:prof}. 
For some pulsars, the scattering time is greater than the rotational period and the pulsations disappear during the folding process. This is the reason why some of the radio sources (J0324$+$5239, B0531$+$21, B0540$+$23, B0611$+$22, B0626$+$24, B1931$+$24, B1946$+$35, B2148$+$63, and B2227$+$61) are not detected: 
their scattering time exceeds the pulsar's period (cf.~Figure~\ref{Scat_P0} and Table~\ref{tab:param:nondetected}). 

For B0355$+$54, the estimated scattering time slightly exceeds the pulse period (Table~\ref{tab:flux}). With this, the pulsar should still remain visible, which is indeed the case (see Figure~\ref{ALL:prof}). Since part of the pulsar's energy reaches the telescope as continuum rather than pulsed emission, the measured flux density only represents the pulsed flux.

We compared the scattering times obtained with YMW16~\citep{yao_new_2017} to those given by a different Galactic density model, namely, NE2001 \citep{cordes_ne2001.i._2002}. The latter model seems to underestimate the scattering with respect to YMW16 
and to the value deduced from our own observed profiles (Figure~\ref{ALL:prof}). This is true in particular for B0355$+$54, B2217$+$47 and B2306$+$55, where the values given by NE2001 are, respectively, 11.6\%, 1.4\%, and 3.4\% of the phase, numbers that are far from those extracted based on our own observed profiles or from the values provided by the electron density model YMW16, namely 108.7\%, 11.0\%, and 16.2\% (cf.~Table~\ref{tab:flux}, column 5).

We note that for some pulsars, the measured scattering index $\alpha_\text{scat}$ (defining the frequency dependence of the scattering time $\tau_\text{scat}$) obtained from observations can deviate considerably from the theoretical value of 4.0 or 4.4 used in models such as in \citet{yao_new_2017}. In particular, \citet{Geyer17} analysed LOFAR observations at 150 MHz and found a less steep behaviour for B0114$+$58, B0540$+$23 and B0611$+$22. If this is confirmed and can be extended to our observing frequency of 60 MHz, the resulting scattering time would be lower and the pulsars would not be rendered undetectable by scattering. In that case, the non-detection of these specific pulsars would be caused by a different reason (see e.g.~Section \ref{sec-disc-nondetections}).

\subsection{Spectral turnover: comparison with HBA census (110-188 MHz)}
\label{sec-comparison-hba}

The 39 pulsars of the FR606 LBA census (25-80 MHz) described in this publication have also been 
detected in the LOFAR HBA census \citep[110-188 MHz]{bilous_lofar_2016}.
The spectral index and turnover frequency given in \citet{bilous_lofar_2016} can be used to estimate a theoretically expected mean flux density for the LBA frequency range and to compare it to our measurements.

\begin{figure*}[ht]
\begin{center}
\includegraphics[scale=0.80]{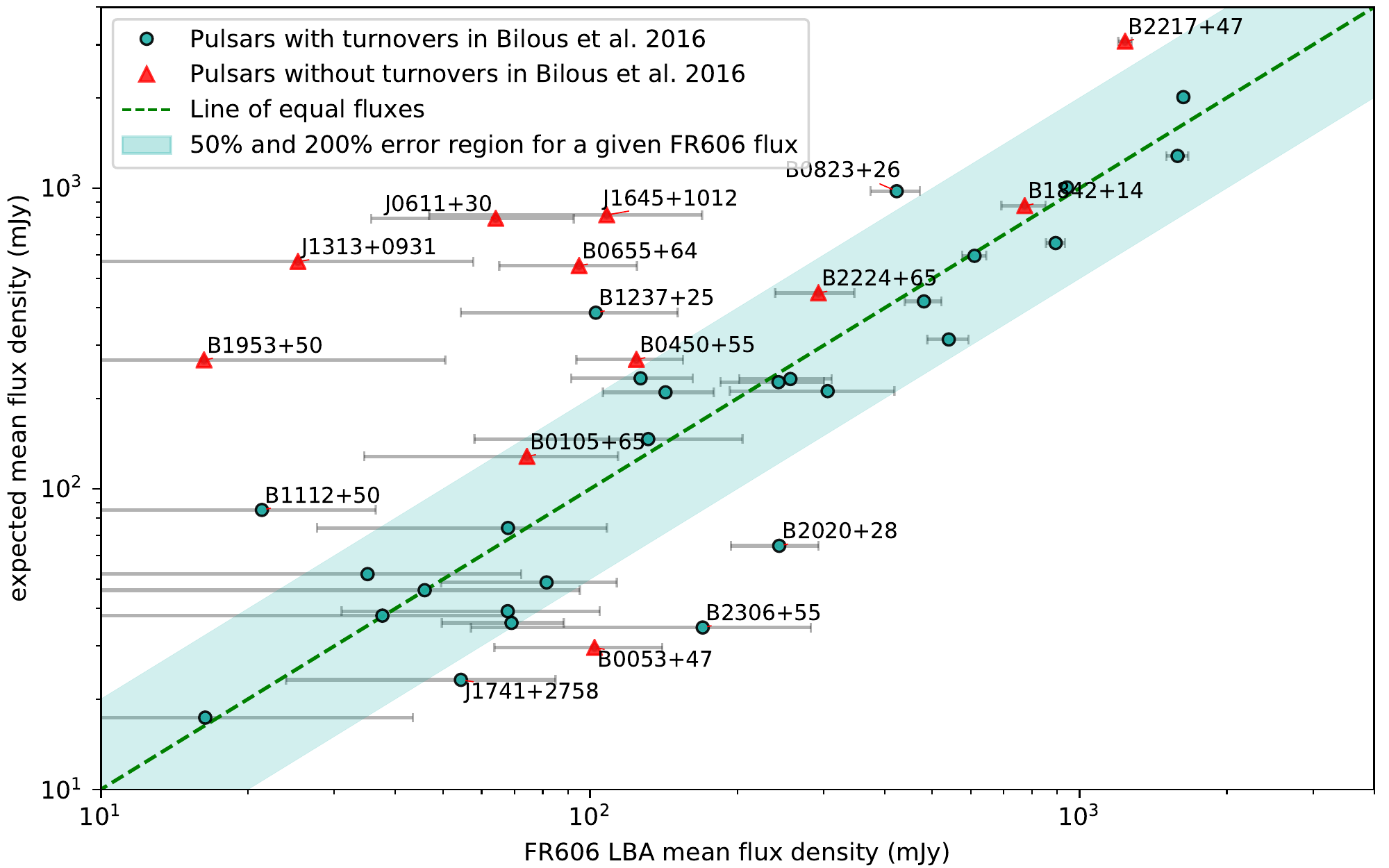}
\end{center}
\caption{Comparison of the mean flux densities reported in this paper with those obtained from the fitted spectral indices in \citet{bilous_lofar_2016}. Blue dots: pulsars fitted with at least one turnover in their spectrum. Red triangles: pulsars fitted with a single spectral index.
The line of equal flux values is shown by a green dashed line. In order to maintain symmetry between the two axes, we take a range from 50-200~\% of the equal flux value (green area) for the systematic error. 
}
\label{Flux_HBA_LBA}
\end{figure*}

Figure \ref{Flux_HBA_LBA} compares the mean flux densities obtained 
from the present LBA census (X-axis) 
to the theoretical mean flux density extrapolated from \citet{bilous_lofar_2016} (Y-axis). 
Pulsars are represented with a blue dot if \citet{bilous_lofar_2016} identified a turnover, 
and a red triangle otherwise (i.e.~the spectrum was fitted using only one spectral index).
For FR606 observations, we indicate the nominal error resulting from the flux calibration. The systematic error of 50\% is represented by the green surface around the diagonal line of equal fluxes.

For most pulsars, the measured and the extrapolated mean flux densities agree within the error range.     
The exceptions are the following:\\
For five of the pulsars  (J0611$+$30, B0655$+$64, J1313$+$0931, J1645$+$1012, and     
                B1953$+$50), the mean flux density extrapolated from the HBA range is considerably higher than the flux density we measured in the LBA range.
                For these pulsars, \citet{bilous_lofar_2016} give a simple power-law without turnover. 
                Our observations show a considerable lack of flux density below 100 MHz, indirectly showing that these pulsars do indeed have a spectral turnover at low frequencies.
                Similarly, in the companion article, \citet{bilous_lofar_2019} find that 
                a low-frequency turnover is compatible with the flux density measurements of J0611$+$30, J1313$+$0931 and J1645$+$1012.
                Similarly, but with a lower significance, we see a hint for a turnover for B2217$+$47.\\  
For B1112$+$50, the extrapolated flux is overestimated with             respect to the measurement. It is, however, consistent with the HBA 
                flux measured in in \citet[][see their Figure C.6]{bilous_lofar_2016}.
                The extrapolation takes this HBA flux into account, but also (older) literature values, suggesting possible time variability of this pulsar.\\
For B1237$+$25, the expected mean flux density is                       overestimated even though \citet{bilous_lofar_2016} fit a spectral                 turnover.
                The model uses three frequency ranges with different spectral indices.
                We suspect that the turnover happens at slightly higher frequency than the 45 MHz estimated in \citet[][see their Figure C.6]{bilous_lofar_2016}.\\
For B2020$+$28 the spectral index of the extrapolation is 
                not well constrained. A shallower spectral index at low frequencies is compatible with both previous observations and our LBA data.\\


This comparison of observations at frequencies below 100 MHz (our work) to observations above 100 MHz \citep{bilous_lofar_2016} shows that several pulsars which used to be described using a single power-law have a spectral turnover.
This does not come as unexpected: \citet{bilous_lofar_2016} found that the average spectral index is lower for measurements at 150 MHz than for higher frequencies (potentially indicating proximity to a turnover) and noted that measurements below 100 MHz are required to study the phenomenon of turnover systematically.

For a number of pulsars which were modelled without spectral turnover \citep{bilous_lofar_2016}, our measured flux density is in agreement with the extrapolated flux density value. This indicates that these pulsars either have no turnover or (more likely) that the turnover occurs at frequencies below our sensitivity maximum ($\sim58$ MHz). 
Again, more high sensitivity observations below 100 MHz are required for a systematic study.

The comparison presented above is just a first step. 
A detailed analysis of spectral indices and turnover frequencies will be presented in a companion publication for the brightest pulsars in our sample (Bondonneau et al. in prep), and more sensitive observations will be provided by NenuFAR in the near future.


\subsection{Comparison with the LOFAR LBA census}

In a companion study, we observed pulsars with the LBA antennas of the LOFAR core \citep{bilous_lofar_2019}. 
Between both studies, there are in total 64 common radio sources. Of these, 36 pulsars were detected by both FR606 and the LOFAR core, 5 were detected only by the LOFAR core, 1 was detected only by FR606, and 22 were not detected by either instrument.

\subsubsection{Common detections}

\begin{figure*}[ht]
\begin{center}
\includegraphics[scale=0.80]{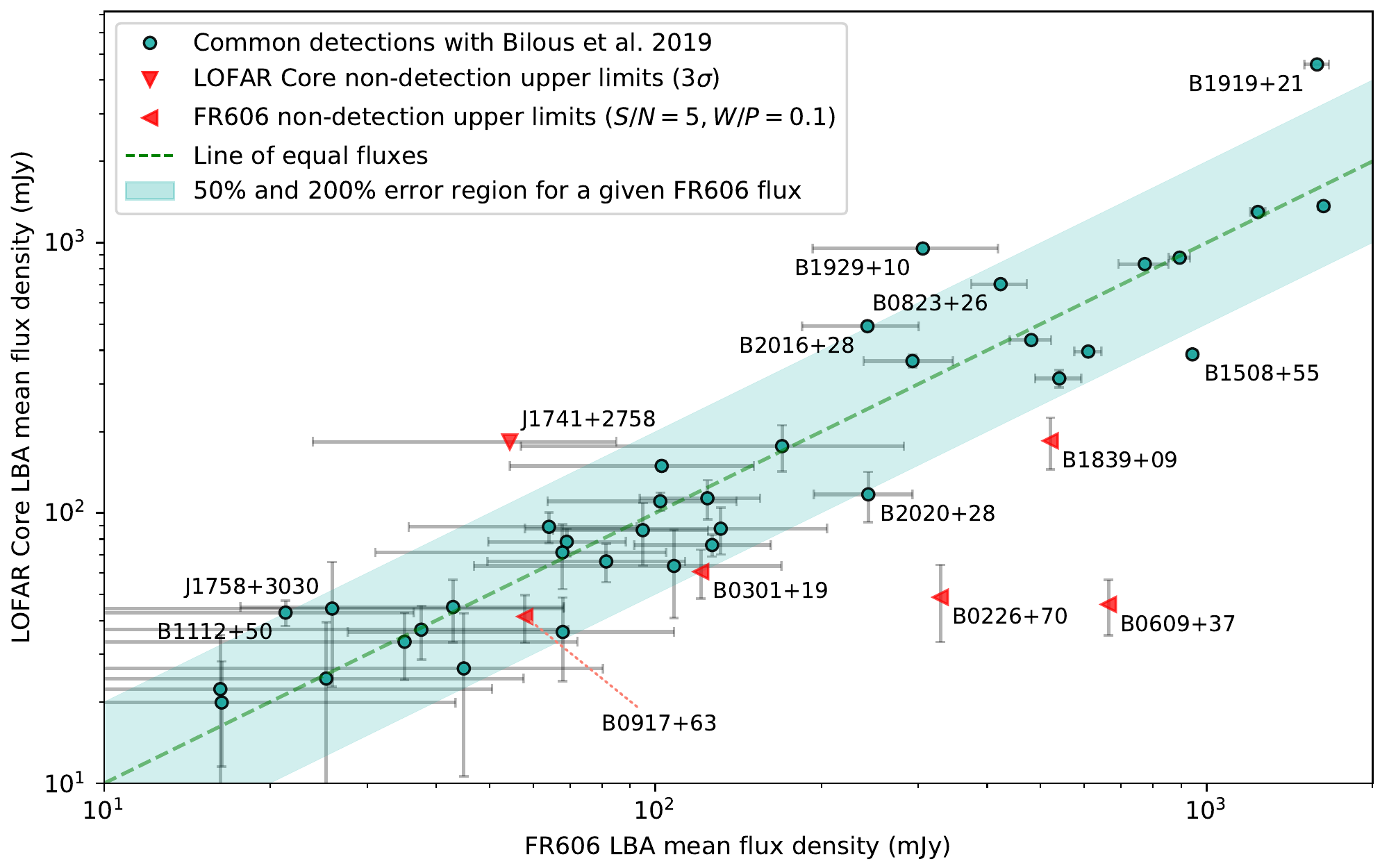}
\end{center}
\caption
{
Comparison of mean flux densities reported in this paper with those obtained by \citet{bilous_lofar_2019}.
The line of equal flux values is shown by green dashed line. In order to maintain symmetry between the two axes, for the systematic error we take a range from 50-200~\% of the equal flux value (green area)
Red triangles show the flux density upper limits
when a pulsar was detected only by FR606 LBA (triangle is pointing down), or only by LOFAR Core (triangles are pointing left). Note that
for the upper limits for the LOFAR Core we used the value of $3\sigma$, where $\sigma$ is the nominal uncertainty on the flux following
\cite{bilous_lofar_2016}, while for the FR606 upper limits, we used Equation~(\ref{equLIM}) from the present paper.}

\label{fig:vlad}
\end{figure*}

Figure \ref{fig:vlad} shows the measured flux densities from the LOFAR Core LBA census (Y-axis) reported by \citet{bilous_lofar_2019}
in comparison with the flux density measurements from the current FR606 census (X-axis). 
For FR606 and LOFAR Core observations, we indicate the nominal error resulting from the flux calibration. The systematic error of 50\% is represented by the green surface around the diagonal dashed line of equal fluxes.

For all of the 36 pulsars that were detected in both censuses, the measured flux densities are compatible or almost compatible within the uncertainties.

Some of the values are not perfectly matched (e.g. B1508$+$55, B1919$+$21, B1929$+$10). 
This can be attributed to a number of reasons \citep[see, e.g.,][]{kondratiev_lofar_2016},
such as the contribution of strong background sources to the wide low-frequency beam, beam jitter caused by the ionosphere, refractive scintillation (RISS), or intrinsic variability. Each of these factors can increase or decrease the measured flux density. 
For instance, both the censuses were performed at different epochs so that RISS could affect the measurements differently. 
For the pulsars B1508$+$55 and B2020$+$28, FR606 has measured a slightly higher flux density than the LOFAR core. This could be explained by ionosphere jitter during the LOFAR Core census observations (the field of view of the international station is wider, so that a small shift of the beam should not matter for our FR606 observations). 
Indeed, \citet{bilous_lofar_2019} used multiple beams simultaneously and found a higher flux density in one of the off-centre beams for several pulsars (including B1508$+$55).

The same factors could potentially lead to non-detections by one of the telescopes or by both. Figure \ref{fig:vlad}  includes pulsars detected by at least one of the telescopes. In case of non-detection by one
of the telescopes, we use the upper limit value on flux density.

\subsubsection{Pulsars detected only by LOFAR core}

Five of the pulsars seen with the LOFAR Core have not been detected by FR606, 
namely PSRs B0226$+$70, B0301$+$19, B0609$+$37, B0917$+$63, and B1839$+$09. 
For all of these non-detections, the upper limits deduced from our FR606 observations are compatible with the measured flux densites of the LOFAR core (see Figure \ref{fig:vlad}).
The non-detection of B0917$+$63 by FR606 in 275 minutes came as a bit of a surprise. It is possible that the RISS leads to intensity variations.
Still, the upper limit of FR606 is compatible with the measured flux density from the LOFAR core.

\subsubsection{Pulsars detected only by FR606}


PSR J1741$+$2758 was not detected with the LOFAR Core in 23 minutes, whereas it was detected by FR606 in 210 minutes. The smaller effective area of FR606 (96 dipoles of FR606 vs. 24x48 dipoles of the Core) is balanced out by the longer integration time.
Also, for the LOFAR Core observation, this particular dataset was of poor quality \citep[more than half of the band was deleted due to dropped packets; see][]{bilous_lofar_2019}.

The non-detection by the LOFAR Core yields an upper flux density limit which is compatible with the detection by FR606.
We also note that this pulsar had already been detected at frequencies below 100 MHz before \citep{zakharenko_detection_2013}.




\subsection{Millisecond pulsars}

Currently, radio detections at frequencies below 100 MHz have been published for four millisecond pulsars
\citep{dowell_detection_2013,kondratiev_lofar_2016,bhat_observations_2018}, of which three are observable by FR606. We have observed and detected all three of these pulsars. 
In view of the low flux densities of these pulsars, we did not include any other millisecond pulsars in our sample.



\subsection{Possible reasons for non-detections}
\label{sec-disc-nondetections}

There are a number of potential reasons for non-detections:\\
The spectrum (as characterised by the spectral index $+$         
        turnover) is not favourable for very low frequency observations.\\
In principle, the pulse period or DM could be outside the range of values probed by \texttt{pdmp}.
        However, all of our pulsars have been previously detected below 200~MHz, so we expect the range in DM to be large enough.
        As we have used updated ephemeris files, we also expect the range in pulse period to be sufficient.\\
        %
The pulse is smeared by scatter-broadening. This is the case, 
        for example, for the Crab pulsar B0531$+$21, where the scattering time is about 500\% of the pulse period. For a number of the pulsars in Table~\ref{tab:param:nondetected}, the expected scatter broadening is high ($\tau_\text{scat}^\text{calc}/P0>1$), which is indeed compatible with our non-detection. See Section \ref{sec:disc:scattering} for details.\\
Intermittent emission such as nulling or mode-changing can affect a pulsars detectability. For example, B0943$+$10 \citep{hermsen13,bilous_lofar_2014}
        and B0823$+$26 \citep{Sobey15}
        are known for their mode-changing behaviour at low frequencies. For mode-changing pulsars, the mean flux density 
        depends on the state of the pulsar during the observation, 
        which can make the difference between a detection and a non-detection. The same is, of course, true for nulling pulsars.\\
Diffractive scintillation should not affect our measurements 
        because the decorrelation bandwidths should be lower than our bandwidth and, thus, many scintles are averaged out.\\
Slow fluctuations of the pulsar amplitudes can be caused by 
        refractive scintillation by the interstellar medium. 
        For observations at 74 MHz, \citet{Gupta93} 
        measured modulation indices (ratio of the standard deviation
        of the observed flux densities to their mean) in the range of
        0.15-0.45, which can account for some of our non-detections.
        The bandwidth they used was of 500 kHz, which is much lower than our bandwidth. However, refractive  scintillation  is  broadband  in nature \citep{Narayan92} so bandwidth should not matter.\\
Some of the flux density values given in earlier publications 
        can be over-estimations, especially for the cases with a low S/N.\\

\section{Conclusion}

In this publication, we observed a total of 102 pulsars, of which 64 were detected successfully. 
Two of these had never before been detected at frequencies below 100 MHz.

We obtained results similar to those in the companion study using the LOFAR core \citep{bilous_lofar_2019}. We were able to partially compensate for the lower effective area ($\sim$10\%) with longer integrations 
during RFI-quiet moments (thus optimising the quality of the data). 
Due to the lower sensitivity of FR606, we did not detect all the pulsars detected by \citet{bilous_lofar_2019} but our upper limits are compatible with their flux density measurements. 
We detected two pulsars that were not part of the sample of \citet{bilous_lofar_2019}, and one pulsar (J1741$+$2758) that was a non-detection in that study.

For several pulsars (J0611$+$30, B0655$+$64, J1313$+$0931, J1645$+$1012, and B1953$+$50), the comparison to observations at slightly higher frequencies \citep{bilous_lofar_2016} indicates a previously unknown spectral turnover.  This confirms the expectation that spectral turnovers are a widespread phenomenon and that measurements below 100 MHz are essential to studying this phenomenon systematically.

We should note that the pulsar population represented in this census is biased by the selection method, essentially based on the previous detections of~\citep{stovall_pulsar_2015, pilia_wide-band_2016, kondratiev_lofar_2016, bilous_lofar_2016}.
It does not take into account pulsars that have not been detected in the HBA range.

In order to further study the population statistics of these low-frequency pulsars, a more homogeneous and substantial dataset is required. 
This will be reached by the NenuFAR radio telescope \citep{ZarkaNenufarSF2A,nenufar,ZarkaNenufarIEEE}
and its pulsar instrumentation LUPPI \citep{BondonneauURSI}, which we are currently using to conduct a systematic census of the known pulsar population (Bondonneau et al. in prep).
NenuFAR, while providing us with an equivalent sensitivity to the LOFAR core at 60 MHz, offers a flat gain response across the LBA frequency band (from 10-85 MHz).
Consequently, a much higher detection rate can be expected than for the present census.
In addition, the flat frequency response will allow a much higher sensitivity towards frequency-dependent effects such as dispersion, scattering, spectral turnovers, and pulsar profile evolution.


\begin{acknowledgements}
This works makes extensive use of \texttt{matplotlib}\footnote{\texttt{https://matplotlib.org}} \citep{Hunter_2007}, \texttt{seaborn}\footnote{\texttt{http://stanford.edu/$\sim$mwaskom/software/seaborn}} Python plotting libraries and NASA’s Astrophysics Data System.

LOFAR, the Low-Frequency Array designed and constructed by ASTRON, has facilities in several countries, that are owned by various parties (each with their own funding sources), and that are collectively operated by the International LOFAR Telescope (ILT) foundation under a joint scientific policy.

The Nan\c{c}ay Radio Observatory is operated by Paris Observatory, associated with the French Centre National de la Recherche Scientifique and Universit\'{e} d'Orl\'{e}ans.

This work was supported by the "Entretiens sur les pulsars" funded by Programme National High Energies (PNHE) of CNRS/INSU with INP and IN2P3, co-funded by CEA and CNES.

The Lovell Telescope is owned and operated by the University of Manchester as part of the Jodrell Bank Centre for Astrophysics with support from the Science and Technology Facilities Council of the United Kingdom. 

The Nançay Radio Observatory is operated by the Paris Observatory, associated with the French Centre National de la Recherche Scientifique (CNRS).

The authors would like to thanks D. Smith for fruitful discussions.

\end{acknowledgements}

\bibliographystyle{aa}

\onecolumn
\begin{appendix}
\section{Non-detection Table}


\addtocounter{table}{-1}
\setlength{\tabcolsep}{9pt}
\topcaption{top}
\tablecaption{Pulsars that were not detected in this census.
JName, BName: pulsar name. P0: pulsar period. 
DM: the DM used to coherently disperse the observations from \citet{zakharenko_detection_2013} \citet{stovall_pulsar_2015} and \citet{bilous_lofar_2016}, and ATNF to complete. $\tau^\text{calc}_\text{scat}$/P0 : the scattering time (estimated using YMW16~\citet{yao_new_2017} at 60 MHz) divided by the pulsar period. 
duration: total duration of the observation in minutes.
elev: the average elevation of the observation.
upper limit for the mean flux: The upper limit for the mean flux density~\ref{limit_flux}.
$^{\eta}$: excluding the contribution of the nebula to $T_\text{sky}$.\
$^{\tau}$: upper limit for the mean flux density is not valid ($\tau^\text{calc}_\text{scat}/$P0>~100~\%).
$^{\epsilon}$: the file is folded using an ephemeris file from either Jodrell Bank Observatory or Nan\c{c}ay Radio Observatory (see Section \ref{fine-tuning}).
\label{tab:param:nondetected}}
\tablefirsthead{\toprule J2000 &Discovery  & P0   & DM            & $\tau^\text{calc}_\text{scat}$/P0 & duration & elev & upper limit for the\\
   Name    &   Name    & sec  & pc.cm$^{-3}$  &   \%   & min    & degree &  mean flux density [mJy]\\    \midrule}
\tablehead{%
\multicolumn{3}{c}%
{{\bfseries  Continued from previous page}} \\
\toprule
J2000 &Discovery  & P0   & DM            & $\tau^\text{calc}_\text{scat}$/P0 & length & elev & upper limit for the\\
   Name    &    Name   & sec  & pc.cm$^{-3}$  &   \%   & min    & degree &  mean flux density [mJy]\\ \midrule}
\tabletail{%
\midrule \multicolumn{3}{r}{{Continued on next page}} \\ \midrule}
\tablelasttail{%
\\\midrule
\multicolumn{3}{r}{{}} \\ \bottomrule}
\begin{xtabular*}{\textwidth}{lllllllll}
J0117$+$5914 & B0114$+$58 & 0.1014 & 49.4230 & 95.3 & 255 &  64 &  344$^{\tau}$ & \\
J0139$+$5814 & B0136$+$57 & 0.2725$^{\epsilon}$ & 73.7790 & 172.6 & 225 &  63 &  366$^{\tau}$ & \\
J0231$+$7026 & B0226$+$70 & 1.4668$^{\epsilon}$ & 46.6400 & 5.3 & 240 &  45 &  329 & \\
J0304$+$1932 & B0301$+$19 & 1.3876$^{\epsilon}$ & 15.7370 & 0.1 & 240 &  53 &  121 & \\
J0324$+$5239 & J0324$+$5239 & 0.3366 & 119.0000 & 984.6 & 240 &  56 &  281$^{\tau}$ & \\
J0415$+$6954 & B0410$+$69 & 0.3907$^{\epsilon}$ & 27.4650 & 2.9 & 360 &  44 &  253 & \\
J0534$+$2200 & B0531$+$21 & 0.0334 & 56.7875 & 497.4 & 60 &  43 &  705$^{\eta,\tau}$& \\
J0543$+$2329 & B0540$+$23 & 0.2460$^{\epsilon}$ & 77.7115 & 235.8 & 120 &  61$^{\tau}$ &  428$^{\tau}$ & \\
J0612$+$3721 & B0609$+$37 & 0.2980$^{\epsilon}$ & 27.1350 & 3.7 & 110 &  42 &  664 & \\
J0614$+$2229 & B0611$+$22 & 0.3350$^{\epsilon}$ & 96.9100 & 426.0 & 240 &  61 &  337$^{\tau}$ & \\
J0629$+$2415 & B0626$+$24 & 0.4766$^{\epsilon}$ & 84.1950 & 168.4 & 180 &  48 &  540$^{\tau}$ & \\
J0653$+$8051 & B0643$+$80 & 1.2144$^{\epsilon}$ & 33.3320 & 1.8 & 240 &  56 &  316 & \\
J0659$+$1414 & B0656$+$14 & 0.3849$^{\epsilon}$ & 13.9770 & 0.4 & 240 &  53 &  77 & \\
J0921$+$6254 & B0917$+$63 & 1.5680$^{\epsilon}$ & 13.1580 & 0.1 & 275 &  51 &  58 & \\
J0943$+$1631 & B0940$+$16 & 1.0874$^{\epsilon}$ & 20.3200 & 0.4 & 115 &  58 &  430 & \\
J0943$+$22 & J0943$+$22 & 0.5329 & 25.1000 & 1.6 & 360 &  65 &  24 & \\
J0947$+$27 & J0947$+$27 & 0.8510$^{\epsilon}$ & 29.0000 & 1.6 & 220 &  65 &  1344 & \\
J1503$+$2111 & J1503$+$2111 & 3.3140 & 3.2600 & 0.0 & 360 &  48 &  57 & \\
J1612$+$2008 & J1612$+$2008 & 0.4266 & 19.5440 & 0.9 & 240 &  56 &  298 & \\
J1627$+$1419 & J1627$+$1419 & 0.4909 & 33.8000 & 4.8 & 180 &  55 &  415 & \\
J1649$+$2533 & J1649$+$2533 & 1.0153$^{\epsilon}$ & 35.5000 & 2.8 & 240 &  64 &  303 & \\
J1720$+$2150 & J1720$+$2150 & 1.6157 & 41.1000 & 3.0 & 240 &  57 &  292 & \\
J1740$+$1000 & J1740$+$1000 & 0.1541$^{\epsilon}$ & 23.8500 & 4.6 & 120 &  43 &  501 & \\
J1752$-$2806 & B1749$-$28 & 0.5626 & 50.3720 & 18.5 & 60 &  14 &  4533 & \\
J1841$+$0912 & B1839$+$09 & 0.3813$^{\epsilon}$ & 49.1070 & 24.8 & 120 &  50 &  521 & \\
J1851$-$0053 & J1851$-$0053 & 1.4091 & 24.0000 & 0.5 & 240 &  38 &  578 & \\
J1907$+$4002 & B1905$+$39 & 1.2358$^{\epsilon}$ & 30.9600 & 1.4 & 250 &  53 &  262 & \\
J1908$+$0734 & J1908$+$0734 & 0.2124 & 11.1040 & 0.4 & 360 &  45 &  203 & \\
J1933$+$2421 & B1931$+$24 & 0.8137 & 105.9251 & 252.3 & 120 &  64 &  468$^{\tau}$ & \\
J1946$+$1805 & B1944$+$17 & 0.4406$^{\epsilon}$ & 16.2200 & 0.5 & 120 &  59 &  110 & \\
J1948$+$3540 & B1946$+$35 & 0.7173 & 129.0750 & 646.7 & 120 &  77 &  391$^{\tau}$ & \\
J1954$+$2923 & B1952$+$29 & 0.4267$^{\epsilon}$ & 7.9320 & 0.1 & 115 &  54 &  124 & \\
J2043$+$2740 & J2043$+$2740 & 0.0961$^{\epsilon}$ & 21.0000 & 4.9 & 115 &  56 &  425 & \\
J2055$+$2209 & B2053$+$21 & 0.8152$^{\epsilon}$ & 36.3610 & 3.8 & 120 &  64 &  419 & \\
J2139$+$2242 & J2139$+$2242 & 1.0835 & 44.1000 & 5.8 & 115 &  57 &  427 & \\
J2149$+$6329 & B2148$+$63 & 0.3801$^{\epsilon}$ & 128.0000 & 1178.8 & 120 &  72 &  1798$^{\tau}$ & \\
J2157$+$4017 & B2154$+$40 & 1.5253$^{\epsilon}$ & 70.8570 & 26.2 & 180 &  36 &  399 & \\
J2229$+$6205 & B2227$+$61 & 0.4431 & 124.6140 & 905.6 & 180 &  47 &  400$^{\tau}$ & \\
\end{xtabular*}

\section{Pulsar profiles}
\begin{figure}[ht]
\begin{center}
\includegraphics[scale=0.75, trim=0cm 10cm 0cm 0cm]{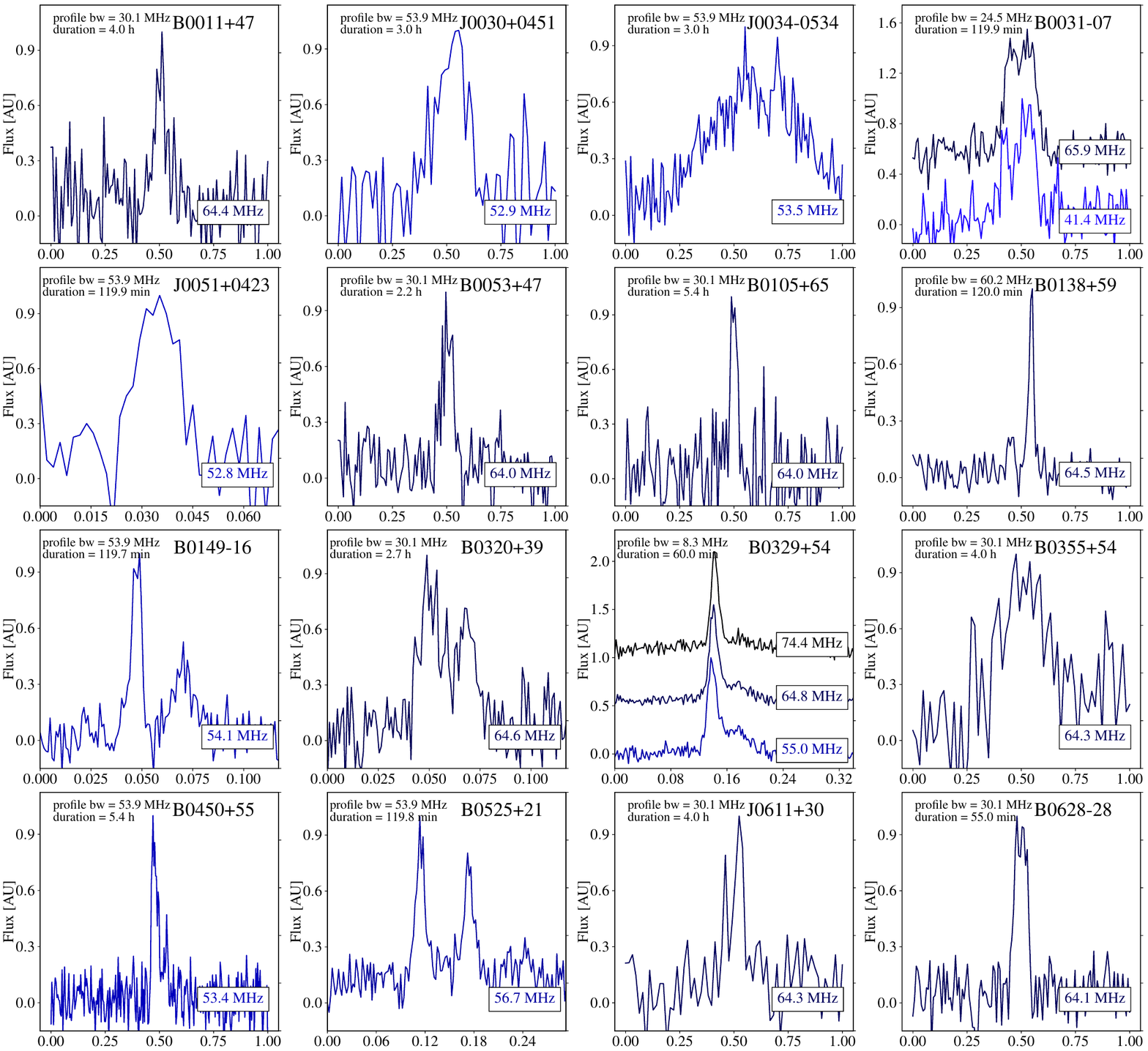}
\end{center}
\end{figure}

\begin{figure}[ht]
\begin{center}
\includegraphics[scale=0.75]{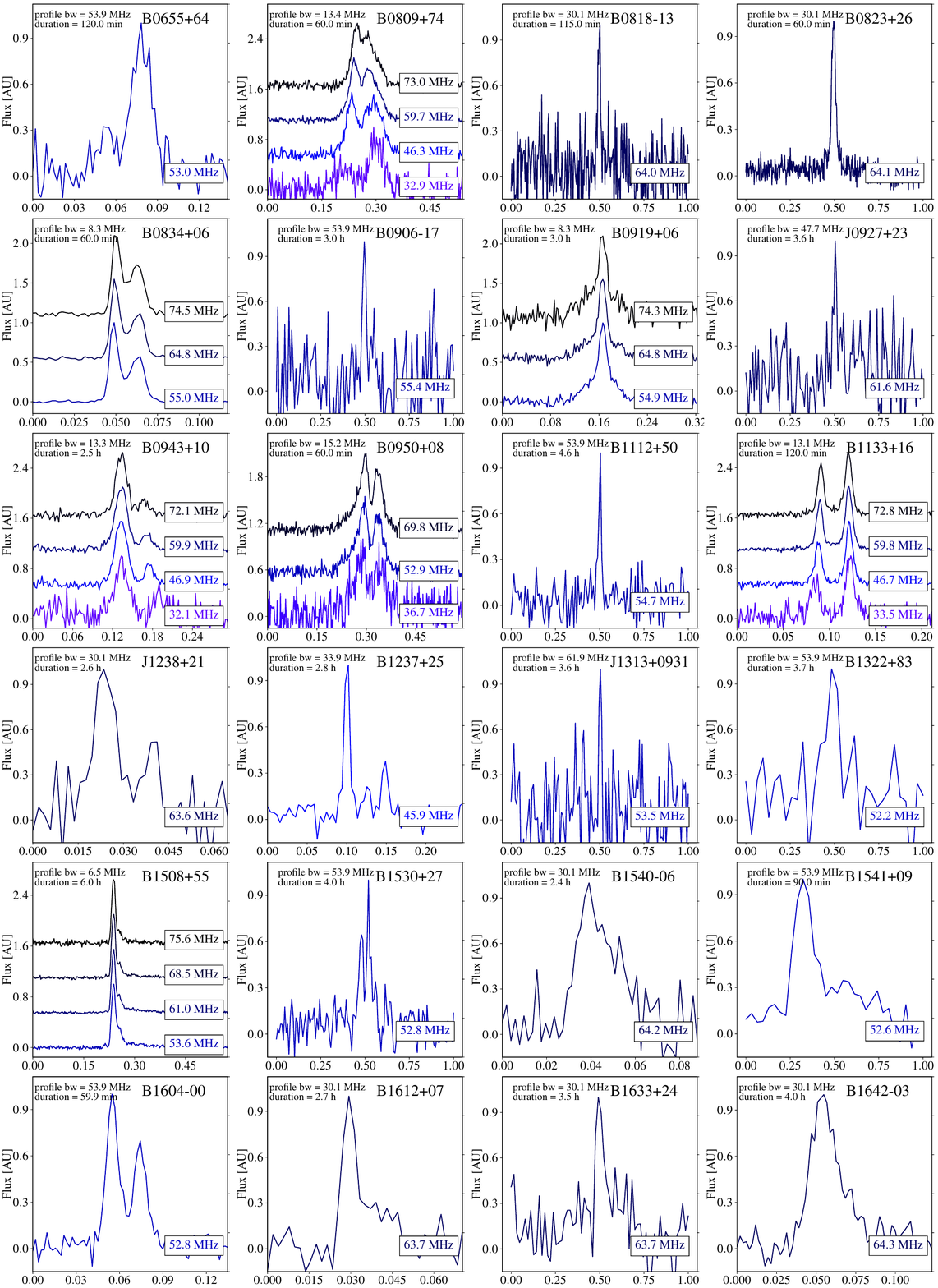}
\end{center}
\end{figure}

\begin{figure}[ht]
\begin{center}
\includegraphics[scale=0.75]{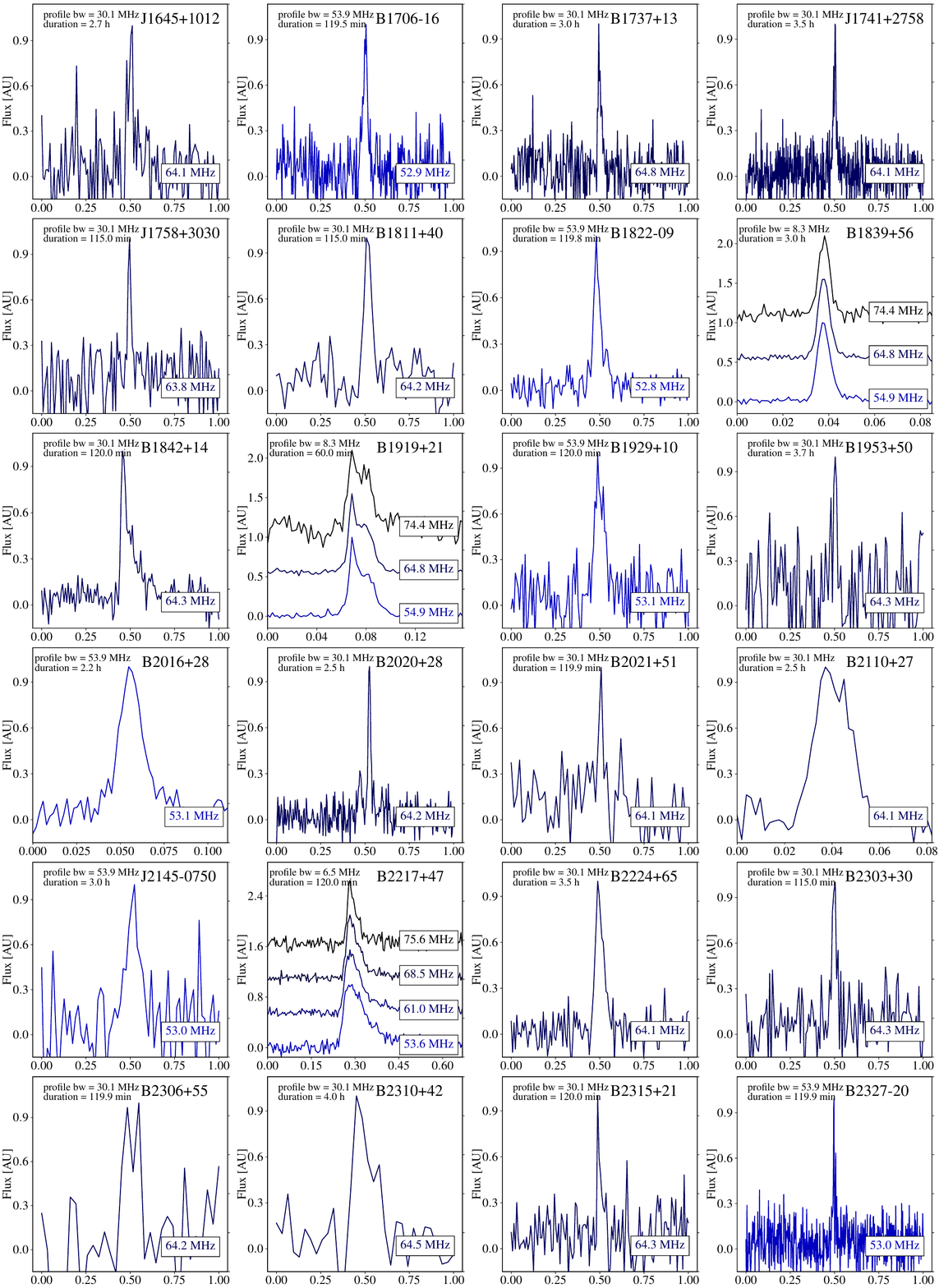}
\end{center}
\caption
{Profiles of the 64 pulsars detected in this study. The profiles are centred on the pulse region. Pulsars with a high S/N are divided into several frequency bands to show frequency-dependent variations in the profiles. At the top left of each sub-figure, the bandwidth and the integration time used for each profile are indicated.}
\label{ALL:prof}
\end{figure}

\end{appendix}

\end{document}